\documentclass[aip]{revtex4-1}

\usepackage{graphicx}
\usepackage{epstopdf}
\usepackage{xcolor}

\usepackage{amssymb}
\usepackage[centertags]{amsmath}
\usepackage{amsthm}
\usepackage[mathcal]{euscript}
\usepackage{tabularx}
\usepackage{fancybox}
\usepackage{wrapfig}

\newcommand{\pD}[2]{\frac{\partial #1}{\partial #2}}

\begin{document}


\title{High-order two-fluid plasma solver for direct numerical simulations of plasma flows with full transport phenomena} 



\author{Z. Li}%
\email{zhaorui.li@tamucc.edu}
\affiliation{ 
Department of Engineering, Texas A$\&$M University-Corpus Christi, Corpus Christi, Texas 78412, USA 
}

\author{D. Livescu}%
\email{livescu@lanl.gov.}
\affiliation{ 
CCS-2, Los Alamos National Laboratory, Los Alamos, New Mexico 87545, USA
}


\begin{abstract}
The two-fluid plasma equations for a single ion species, with full transport terms, including temperature and magnetic field dependent ion and electron viscous stresses and heat fluxes, frictional drag force, and ohmic heating term have been implemented in the CFDNS code and solved by using sixth-order non-dissipative compact finite differences for plasma flows in several different regimes. In order to be able to fully resolve all the dynamically relevant time and length scales, while maintaining computational feasibility, the assumptions of infinite speed of light and negligible electron inertia have been made. Non-dimensional analysis of the two-fluid plasma equations shows that, by varying the characteristic/background number density, length scale, temperature, and magnetic strength, the corresponding Hall, resistive, and ideal magnetohydrodynamics (MHD) equations can be recovered as limiting cases. The accuracy and robustness of this two-fluid plasma solver in handling plasma flows in different regimes have been validated against four canonical problems: Alfven and whistler dispersion relations, electromagnetic plasma shock,  and magnetic reconnection. For all test cases, by using physical dissipation and diffusion, with negligible numerical dissipation/diffusion, fully converged Direct Numerical Simulations (DNS)-like solutions are obtained when the ion Reynolds number based on grid size is smaller than a threshold value which is about $2.3$ in this study. For the magnetic reconnection problem, the results show that the magnetic flux saturation time and value converge when the ion and magnetic Reynolds numbers are large enough. Thus, the DNS-like results become relevant to practical problems with much larger Reynolds numbers.
\end{abstract}


\maketitle

\section{Introduction}
\label{sec:1}

Plasma, by far the most abundant form of ordinary matter in the universe, has been the subject of study in many disciplines, particularly in fusion~\cite{Chen1984,Nakai1996,Pfalzner2006}, space physics~\cite{Priest1984,Kallenrode1998,Birn2001}, industrial applications~\cite{Boulos1991,Jeong1998,Gomez2009}, astrophysics~\cite{Kaplan1973,Orlando2007,Zweibel2009}, and so on. Aside from full {\em ab initio} descriptions~\cite{Kohn1965,Hu2016,Ding2018}, for many applications, a reasonable level of accuracy for plasma flows calculations can be achieved by using kinetic theory and the distribution functions that characterize each particle component~\cite{Montgomery1964}. The evolution of the distribution functions is governed by the Boltzmann equation~\cite{braginskii1965}. However, for turbulent flows, solving the six-dimensional Boltzmann equation coupled with Maxwell’s equations for the electromagnetic field is prohibitively expensive, due to the broad range of scales that need to be captured. Using the continuum approximation, when possible, becomes computationally necessary for the description of turbulent plasma flows because the governing equations solved in the fluid model are three-dimensional. Assuming quasi-local thermal equilibrium (i.e. small departures from Maxwellian distribution function) within each of the components, the fluid equations describing plasma dynamics can be obtained by taking appropriate moments of Boltzmann equation and averaging over velocity space for each of the components~\cite{Chapman1970,Cercignani1988}. 

For single component plasmas, i.e. consisting of electrons and a single ion component, starting from the equations for the ion and electron distribution functions, Braginskii ~\cite{braginskii1965} derived a two-fluid hydrodynamic model for separate ion and electron fluids by using the Chapman-Enskog expansion with two-term Sonine polynomial solutions. In the Braginskii two-fluid model, the transport terms include the magnetic field impact on viscous stress tensor, heat flux, and frictional drag force, with different formulations along and perpendicular to the magnetic field. In contrast to single ion component case, plasma equations containing multiple ion species involve additional transport phenomena such as baro- and electro-diffusion~\cite{Amendt2011,Kagan2012,Yin2016}. Although this is a very active area of research~\cite{Simakov2014,Simakov2016,Vold2017,Vold2018}, there are still many open questions, especially on how to treat mixtures with magnetic field dependent transport properties.

According to the H-theorem of Boltzmann~\cite{Chapman1970,Sommerfeld1964}, if the distribution function changes only by virtue of collisions, any arbitrary distribution will approach a Maxwellian. Therefore, the Braginskii two-fluid plasma model~\cite{braginskii1965} can describe well plasma flows in which the characteristic time scale is much larger than the collision time, i.e. $t_{0}\gg\tau_{s}$, and the characteristic length scale is much larger than the distance traversed by particles between collisions (e.g. particle mean-free-path), i.e. $L_{0}\gg\lambda_{\text{mfp}}$. One of such applications is the study of hydrodynamic instabilities between the hot spot and the colder surrounding plasma during the Inertial Confinement Fusion (ICF) coasting/deceleration stage . For the DT plasma in the early deceleration stage, the primary parameters of interests found in the literature~\cite{GL16,Weber2014}, i.e. reference number density $n_{0}\sim\text{10}^{\text{30}}\text{m}^{\text{-3}}$, temperature $T_{0}\sim\text{2.5keV}$, acceleration $g\sim1.0\times\text{10}^{\text{14}}m/s^{2}$, and hot-spot radius $R_{hs}\approx55\mu m$, lead to $\tau_{RT}/\tau_{i}\approx250\gg1$ and $R_{hs}/\lambda_{\text{mfp}}\approx150\gg1$. Here, $\tau_{RT}=\sqrt{1/{(Akg)}}$ is the classical single-mode Rayleigh-Taylor instability (RTI) growth time and $\tau_{i}$ is the ion collision time. The definitions of Atwood number, $A$, and wavenumber, $k$, can be found in Ref.~\cite{WL12}. Similarly, by using the typical plasma scales for the late deceleration stage~\cite{Betti2001,Edwards2011}, one obtains $\tau_{RT}/\tau_{i}\approx500\gg1$ and $R_{hs}/\lambda_{\text{mfp}}\approx320\gg1$.

Unfortunately, magnetized plasmas encountered in nature, including space and astrophysical plasmas, are mostly collisionless, and the typical collision time and mean-free-path in such flows can be comparable to or even larger than certain characteristic time and length scales of the flow. For example, according to the primary parameters given in Refs.~\cite{Parker1983,Browning2014}, the particle mean-free-path in solar flare/corona is much larger than the length-scale of reconnecting current sheet, i.e. $\lambda_{\text{mfp}}\approx1.0\times\text{10}^{\text{5}}\text{m}\gg\delta\sim\text{10m}$. Therefore, the quasi-Maxwellian distribution (or quasi-local thermal equilibrium) assumption does not seem guaranteed in such regimes.  Theoretically, the highly collisionless  magnetic reconnection can only be rigorously described by using collisionless kinetic models (e.g. Vlasov-Maxwell system of equations) in which both ion and electron kinetic-scale features are included~\cite{Daughton2011}. However, in fact, the fluid model can still describe fairly well some strongly magnetized, collisionless plasma dynamics~\cite{Chen1984,Fitzpatrick2014}, which is largely due to the following two justifications. First, a strong magnetic field can play the role of collisions by forcing particles to gyrate in a Larmor orbit that is smaller than the mean free path by a factor of $\omega_{ce}\tau_{e}$~\cite{braginskii1965,Fitzpatrick2014}, where $\omega_{ce}$ is electron cyclotron frequency and $\tau_{e}$ is the electron collision time~\cite{Chen1984}. The other argument is that, even though the real distribution function in collisionless plasmas may significantly deviate from Maxwellian distribution, the fluid equations derived based on the quasi-Maxwellian assumption may approach the physical solution when the range of fluid scales is very broad. This argument is similar to the mixing transition~\cite{Dimotakis00} often invoked in fluid turbulence to justify the relevance of finite Reynolds number simulations to practical problems with much larger range of scales~\cite{CCM04}.
A more rigorous statement of the argument is that the flow develops an inertial range, where the energy cascade is local~\cite{Eyink05} and not influenced by the viscosity, except through the magnitude of the mean dissipation. From this point of view, it is tempting to assume that numerical dissipation in Euler equations simulations can act in a  similar way and allow the development of an inertial range, so that the numerical solution is close to the physical solution when the grid is fine enough. However, developing a power law range in the spectrum is not proof of the emergence of an inertial range~\cite{Zhao_Hussein_2018} and proving cascade locality in the presence of numerical viscosity/diffusion may be impossible in general. In a broader sense, mixing transition may be extended to certain non-turbulent flows to mean convergence of the results with respect to the Reynolds number. In other instances, the concept of separation of scales can also be used to justify the relevance of fluid simulations to practical applications. For example, when the shock wave thickness is much smaller than the flows scales, the results become independent of the shock profile. In this case, even though the Navier-Stokes description breaks around strong shocks, it can still accurately predict the shock-turbulence interaction~\cite{ryu2014}. 
While the mixing transition has not been explicitly explored for plasma flows,  it has implicitly been assumed  for example by showing that two-fluid plasma (including Hall-MHD) equations can successfully predict the fast reconnection rate in collisionless magnetic reconnection~\cite{Biskamp1997,Birn2001,Ma2001,Shay2001,Hakim2006}. Here, we further address this issue by considering the convergence of magnetic reconnection results as the ion and magnetic Reynolds numbers are increased.

Single-fluid magnetohydrodynamics (MHD) has been successfully used for studying large-scale plasma flows in a wide range of problems~\cite{Freidberg1982,Ogino1986,Gombosi1994,Mikic1999,Groth2000,Chapman2004}. However, single-fluid MHD fails to describe plasma phenomena that happen on a length scale comparable to or smaller than the ion skin depth, i.e. when $L_{0}\leq\lambda_{i}=c/\omega_{pi}$, where $c$ is the speed of light in vacuum and $\omega_{pi}$ is ion plasma frequency. When applied to the magnetic reconnection problem, ideal MHD cannot predict the reconnection due to its flux-frozen-in limitation, while resistive MHD predicts a growth rate much lower than observations~\cite{Birn2001}. This is because two-fluid effects become important at length scales below $\lambda_i$  as the ions and electrons motions start to decouple. By including the Hall current in the governing equations and electron pressure contribution to the total pressure, Hall-MHD equations~\cite{Huba2003} account for some two-fluid effects and have been successful in capturing the rapid magnetic reconnection process~\cite{Birn2001,Shay2001,Ma2001,Otto2001}. Nevertheless, Hall-MHD equations are not as general as the Braginskii two-fluid plasma model. For example, to close the Hall-MHD equations, some studies neglect the electron pressure altogether~\cite{Callen2006,Dmitruk2006,Toth2008}, while others assume identical ion and electron temperatures~\cite{Browning2014,Srinivasan2011}. In addition, when viscous effects were included in Hall-MHD equations~\cite{Dmitruk2006,Browning2014}, again these were less general than those in the Braginskii two-fluid plasma model.

In many practical problems, electron and ion temperatures are  different. For example, the ion temperature in the Saturnian magnetosphere near Titan’s orbit is normally higher than the electron temperature; while in Titan’s ionosphere, the electron temperature is dominant over the ion temperature~\cite{Ma2011}. The two-fluid plasma model~\cite{braginskii1965,Freidberg1982,Shumlak2003,bond2017} can solve many of the problems encountered in single-fluid MHD and Hall-MHD formulations by considering separate ion and electron set of equations. Nevertheless, previous applications of the two-fluid model~\cite{Shumlak2003,Loverich2011,bond2017} did not include plasma transport terms and relied on the numerical dissipation/diffusion to obtain stable solutions; such solutions can obviously become corrupted by numerical artifacts and generally might misrepresent the physical transport. 

Previous studies of plasma flows with physical transport phenomena such as thermal diffusion include simulations with Flash~\cite{FLASH1,FLASH2}, Hydra~\cite{HYDRA1,HYDRA2}, and Miranda~\cite{Miranda1, Weber2014, Weber2015} codes. In these codes, the transport coefficients for thermal diffusion were calculated by using the Lee-More model~\cite{Lee-More1984}. However, Flash and Hydra codes solve the inviscid fluid equations and only Miranda explicitly considers viscous and diffusive effects~\cite{Murillo2008}. In particular, the MIRANDA code uses a very similar high-order numerical scheme as the CFDNS code, with negligible numerical dissipation; however, this is accompanied by a high order filter to remove high frequency oscillations. No filtering is used with the CFDNS code. As far as we know, the magnetic field impact on the transport phenomena perpendicular to the magnetic field has not been considered in previous two-fluid plasma flow simulations. Nevertheless, the presence of a strong magnetic field reduces the distance traveled by particle during collision. As a result, depending on the magnetic strength, the plasma transport coefficients in the directions perpendicular to the magnetic field may become significant small, so that the associated fluxes become strongly anisotropic. As argued above, there are many situations, e.g. when a mixing transition exists, where the exact form of the physical transport is not important, provided that the energy transfer among scales of motion remains local. Nevertheless, such transition and the role of anisotropic transport have not been explored for many of the practical situations of interest.

The objective of this study is to present an accurate two-fluid plasma solver with a single ion component that can simulate magnetized plasma flows in a range of applications, with a special focus on collisional dominated transport for low-Z fully ionized nondegenerate plasmas, in regimes where the results might be sensitive to the exact formulation of the transport terms. All plasma transport terms such as the temperature and magnetic field dependent ion and electron viscous stresses and heat fluxes, frictional drag force, and ohmic heating are included in the two-fluid plasma solver. To obtain fully-resolved Direct Numerical Simulations (DNS)-like solutions, the two-fluid plasma equations are solved by using sixth-order non-dissipative compact finite differences~\cite{Lele92} at sufficiently fine grid resolutions. In this study, to maintain computational feasibility, the infinite speed of light and negligible electron inertia assumptions are made to eliminate severe time-step limitations. These two assumptions can be well justified for problems such as ICF coasting stage, where ion thermal velocity is non-relativistic, $V_{Ti}/c\sim O\left( \text{10}^{\text{-3}}\right)$, and $m_{i}/m_{e}\sim 5\times\text{10}^{\text{3}}$. The length scale limitation imposed by using these two assumptions, $L_{0}\gg\left(r_{Le},\lambda_{e}\right)$, where $r_{Le}$ is electron Larmor radius and $\lambda_e$ is electron skin depth, is also satisfied in many other practical problems. While the primary target applications for the new solver are plasma flows which can be described with collisional transport terms, the test problems considered are widely used in the literature and have been addressed primarily using ideal equations solvers; the numerical treatment of such equations requires numerical dissipation/diffusion for regularization. Our new solver yields smooths solutions without any numerical dissipation/diffusion and can recover inviscid analytical solutions for sufficiently high Reynolds numbers. 

In general, the Braginskii transport coefficients become inaccurate for degenerate partially ionized plasmas or high-Z materials~\cite{Lee-More1984}. However, more general formulations do not include full directional dependence of the physical transport with respect to the magnetic field. A separate objective of this study is to form the basis of future estimations of anisotropic transport importance and explore the existence of a mixing transition in various applications.

The paper is organized as follows: in Section II, the derivations of reduced two-fluid plasma equations from the Braginskii’s full two-fluid plasma model, together with an analysis of their ranges of applicability, are discussed in details. A non-dimensional analysis of the reduced two-fluid plasma equations is conducted in Section III. The accuracy and robustness of the two-fluid plasma solver are highlighted, in Section IV, against a series of canonical problems. Finally, the main conclusions are provided in Section V. 

\section{MATHEMATICAL FORMULATION}
\label{sec:2}

The macroscopic description of plasma in fluid theory can be obtained by taking appropriate moments of Boltzmann equation and averaging over velocity space for each of the components in plasma. When using the Chapman-Enskog expansion~\cite{Chapman1970,Cercignani1988}, the zeroth-order distribution function for each species, $f_{s}^{0}$, is chosen to be a Maxwellian, which assumes a local thermal equilibrium within each of the components. By considering the effects that produce small deviations from equilibrium, Braginskii~\cite{braginskii1965} derived a set of two-fluid plasma transport equations and constitutive relations for all transport terms.  On the other hand, ignoring those effects leads to a set of Euler-type two-fluid plasma equations in which all transport phenomena are absent~\cite{Shumlak2003,bond2017}. Such equations develop singularities in finite time and need to be regularized by the numerical algorithm. 

\subsection{Braginskii's two-fluid plasma model}
\label{sec:twofluid}

For a simple fully ionized plasma, the continuity, momentum, and internal energy transport equations for species $s$ ($s=i$ for ion and $s=e$ for electron) are given below as\cite{braginskii1965}:
\begin{eqnarray}
\label{eq:1}
\pD{\rho_{s}}{t}+\nabla\cdot\left(\rho_{s}\textbf{u}_{s}\right)&=&0,\\
\label{eq:2}
\pD{\left(\rho_{s}\textbf{u}_{s}\right)}{t}+\nabla\cdot\left(\rho_{s}\textbf{u}_{s}\textbf{u}_{s}\right)&=&-\nabla{p_{s}}-\nabla\cdot\mathbf{\pi}_{s}+\frac{q_{s}\rho_{s}}{m_{s}}\left(\textbf{E}+\textbf{u}_{s}\times\textbf{B}\right)+\textbf{R}_{s}+\rho_{s}\textbf{g},\\
\label{eq:3}
\pD{\left(\rho_{s}e_{s}\right)}{t}+\nabla\cdot\left(\rho_{s}e_{s}\textbf{u}_{s}\right)&=&-p_{s}\nabla\cdot\textbf{u}_{s}-\nabla\cdot\textbf{q}_{s}+\mathbf{\pi}_{s}:\nabla\textbf{u}_{s}+Q_{s},
\end{eqnarray}

\noindent
where the primary variables are species density, $\rho_{s}$, velocity, $\textbf{u}_{s}$, and specific internal energy, $e_{s}$. In this study, ideal gas equation of state (EOS) is assumed for simplicity. Therefore, the species pressure can be expressed as $p_{s}=\left(\gamma-1\right)\rho_{s}e_{s}=n_{s}k_{B}T_{s}$, in which $\gamma$ and $k_{B}$ are specific heat and Boltzmann constant, while $n_{s}=\rho_{s}/m_{s}$ and $T_{s}$ are species number density and temperature, respectively. $m_{s}$ and $q_{s}$ are mass and charge of particle $s$. The ion and electron charges are $q_{i}=Ze$ and $q_{e}=-e$, in which $e$ is the constant elementary charge. The formulations for plasma transport terms in the above equations, including species viscous stress, $\mathbf{\pi}_{s}$, heat flux, $\textbf{q}_{s}$, frictional drag force, $\textbf{R}_{s}$, and collision generated heat, $Q_{s}$, can be found in Appendix \ref{sec:A} and Ref. \cite{braginskii1965}. The accuracy of Braginskii transport coeefficients for the domain of applicability has been confirmed by comparing with the transport coefficients predicted by using Ab Initio Molecular Dynamics (AIMD)~\cite{Burakovsky2013,Hu2014}. For example, for the DT hot-spot in ICF with number density $n_{e}\sim\text{10}^{\text{31}}\text{m}^{\text{-3}}$ and temperature $T_{e}\sim\text{10.0 keV}$, the electron thermal conductivity calculated by using Braginskii model (see Appendix \ref{sec:A} for full definitions) is $\kappa_{Te}=n_{e}k_{B}^{2}T_{e}\tau_{e}/m_e=\left(n_{e}k_{B}^{2}T_{e}/m_e\right)\times\left[6\sqrt{2}\pi^{3/2}\varepsilon_{0}^{2}\sqrt{m_{e}}\left(k_{B}T_{e}\right)^{3/2}/\left(ln\Lambda e^{4}Zn_{e}\right)\right]\approx 5.15\times10^{9}\left(\text{W}\text{m}^{-1}\text{K}^{-1} \right)$ which is very close to the AIMD prediction \cite{Hu2014}, i.e. $\kappa_{e}=5.05 \times10^{9}\left(\text{W}\text{m}^{-1}\text{K}^{-1} \right)$.      

As discussed before, the Braginskii two-fluid plasma model is  derived for the collision-dominated plasma flows in which the characteristic time and length scales are much larger than the collision time and particle mean-free-path, i.e. $t_{0}\gg\tau_{s}$ and $L_{0}\gg\lambda_{\text{mfp}}$. In addition, the Brangiskii transport coefficients become inaccurate for degenerate partially ionized or high-Z plasmas~\cite{Lee-More1984}. Thus, the results presented here apply to fully ionized nondegenerate single low-Z ion component collisional plasmas, or plasma flows where a mixing transition has occurred.

The evolutions of electric field, $\textbf{E}$, and magnetic field, $\textbf{B}$, are governed by the Maxwell equations as given below: 
\begin{eqnarray}
\label{eq:4}
\frac{1}{c^2}\pD{\textbf{E}}{t}&=&\nabla\times\textbf{B}-\mu_{0}\textbf{J},\\
\label{eq:5}
\pD{\textbf{B}}{t}&=&-\nabla\times\textbf{E},\\
\label{eq:6}
\nabla\cdot\textbf{E}&=&\dfrac{\rho_{c}}{\varepsilon_{0}},\\
\label{eq:7}
\nabla\cdot\textbf{B}&=&0,
\end{eqnarray}

\noindent
where $\mu_{0}$ and $\varepsilon_{0}$ are the permeability and permittivity of free space, respectively, and are related to the speed of light in vacuum, $c$, as $c=\left(\mu_{0}\varepsilon_{0}\right)^{-1/2}$. In the above equations, the formulations for current density, $\textbf{J}$, and local charge density, $\rho_{c}$, are $\textbf{J}=\sum{q_{s}n_{s}\textbf{u}_{s}}=e\left(Zn_{i}\textbf{u}_{i}-n_{e}\textbf{u}_{e}\right)$ and $\rho_{c}=\sum{q_{s}n_{s}}=e\left(Zn_{i}-n_{e}\right)$, respectively. It is worth pointing out that, for closing the governing equations, one only needs solve two of the Maxwell equations and the other two equations (e.g. Eqs.(\ref{eq:6})-(\ref{eq:7})) are just restatements of the  closed set of governing equations. For example, by multiplying continuity equation (\ref{eq:1}) by $q_{s}/m_{s}$ and then taking summation over ion ($s=i$) and electron ($s=e$) species, one obtains:
\begin{equation}\label{eq:8}
\pD{\rho_{c}}{t}+\nabla\cdot\textbf{J}=0.
\end{equation}

\noindent
Furthermore, by taking the divergence of Ampere’s equation (\ref{eq:4}) and then subtracting it from equation (\ref{eq:8}), it yields: 
\begin{equation}\label{eq:9}
\frac{\partial}{\partial{t}}\left(\nabla\cdot\textbf{E}-\rho_{c}/\varepsilon_{0} \right)=0.
\end{equation}

\noindent
Obviously, the Gauss equation (\ref{eq:6}) is just a restatement of the consequence (i.e. Eq. (\ref{eq:9})) of solving continuity equation (\ref{eq:1}) and Ampere equation (\ref{eq:4}). In this study, we would like to call equation (\ref{eq:6}) a diagnostic equation instead of a redundant equation. This is because that, after applying the two assumptions discussed in the following subsections, the electric field is calculated from the generalized Ohm’s law (\ref{eq:13}) instead of the Ampere equation (\ref{eq:4}), and equation (\ref{eq:8}) is reduced to a quasi-neutrality condition. As a result, equation (\ref{eq:9}) is no longer rigorously guaranteed to be satisfied when solving the final governing equations given in section \ref{sec:finaleq}. Therefore, in this study, we solve equation (\ref{eq:6}) as a diagnostic tool to monitor the importance of the numerical integration errors. By following the same procedure, one can also conclude that, mathematically, the magnetic field remains divergence free if it is initially divergence free.  

\subsection{Infinite speed of light assumption}
\label{sec:light}

In this study, the severe time-step restrictions~\cite{Srinivasan2011} (e.g. $\Delta t \leq \text{CFL}\ \Delta{x}/c$ and $\Delta t \leq0.1/\omega_{pe}$, where $CFL$ is the Courant-Frederic-Levi constant, $\Delta x$ is the mesh size, and $\omega_{pe}$ is electron plasma frequency) caused by high frequency electromagnetic waves are eliminated by using the infinite speed of light assumption, i.e., $\left(\partial{\textbf{E}}/\partial{t}\right)/c^2\approx 0$, which reduces the Ampere’s equation (\ref{eq:4}) to:
\begin{equation}\label{eq:10}
\textbf{J}=\frac{1}{\mu_{0}}\nabla\times\textbf{B}.
\end{equation}

Consequently, this assumption restricts the calculations to plasma flows with nonrelativistic thermal velocity, $V_{Ts}=\sqrt{k_{B}T_{s}/m_{s}}\ll{c}$, and to electromagnetic waves with phase speed, $V_{p}=\omega/k\ll{c}$ (see also Ref. \cite{Freidberg1982}). Furthermore, by replacing equation (\ref{eq:10}) into equation (\ref{eq:8}), one obtains:
\begin{equation}\label{eq:11}
\pD{\rho_{c}}{t}=0,
\end{equation}

\noindent
which indicates that the quasi-neutrality condition ($\rho_{c}=0$) is maintained at all times if the initial plasma flow is charge free. Consistently, the number densities and mass densities of ions and electrons become dependent, i.e. $n_{e}=Zn_{i}$ and $\rho_{e}=Z\left(m_{e}/m_{i}\right)\rho_{i}$, which eliminates the need to solve the continuity equation (\ref{eq:1}) for electrons and relates the ion and electron velocities via the current density as,
\begin{equation}\label{eq:12}
\textbf{u}_{e}=\textbf{u}_{i}-\left(\frac{1}{eZn_{i}}\right)\textbf{J}.
\end{equation}

The quasi-neutrality condition limits our interests to plasma phenomena whose characteristic frequency is much smaller than the electron plasma frequency, $\omega\ll\omega_{pe}=\sqrt{n_{e}e^{2}/\varepsilon_{0}m_{e}}$, and characteristic length is much larger than the Debye length, $L_{0}\gg\lambda_{De}=V_{Te}/\omega_{pe}$ (see also Ref.~\cite{Freidberg1982}). 

\subsection{Negligible electron inertia assumption}
\label{sec:electronmass}

The second assumption made in this study is negligible electron inertia in the electron momentum equation (\ref{eq:2}). This assumption is justified as the right hand side of the electron momentum equation is the same order as that of the ion momentum equation, but the advection part is the order $m_e/m_i$ compared to the corresponding part of the ion momentum equation. Then, after applying the relation between ion and electron velocities (equation \ref{eq:12}), one can obtain the generalized Ohm’s law as:
\begin{equation}\label{eq:13}
\textbf{E}=\left(\frac{m_{i}}{eZ\rho_{i}}\right)\left[-\nabla{p_{e}}-\nabla\cdot\mathbf{\pi}_{e}+\textbf{R}_{e}+Z\left(\frac{m_{e}}{m_{i}}\right)\rho_{i}\textbf{g}+\textbf{J}\times\textbf{B}\right]-\textbf{u}_{i}\times\textbf{B},   
\end{equation}

\noindent
where Biermann battery, viscous, resistive, acceleration, and Hall effects are all included. Recent kinetic simulations~\cite{Amendt2011} show that the Biermann battery term appearing in equation (\ref{eq:13}) is the physical source of strong, self-generated electric fields observed in ICF plasma~\cite{Rygg2008}. The rest of the terms, in particular, the Hall term and the last term in equation (\ref{eq:13}) are also indispensable in maintaining the constant charge condition (i.e. Eq. \ref{eq:11}).

Negligible electron inertia implies that the electron flow has an infinite fast response time on the time scales of interest. Therefore, the characteristic time scale of interest must be larger than electron plasma frequency and electron cyclotron frequency, i.e., $1/\omega\gg\left(1/\omega_{pe},1/\omega_{ce}\right)$, which further relaxes the time-step restriction on $0.1/\omega_{ce}$~\cite{Srinivasan2011}. Consistently, the characteristic length scale of interest must be longer than the Debye length, the electron Larmor radius, and/or electron skin depth, i.e. $L_{0}\gg\left(\lambda_{De},r_{Le}, \text{ and/or } \lambda_{e} \right)$, where $r_{Le}=V_{Te}/\omega_{ce}$, $\lambda_{e}=V_{A}/\omega_{ce}=c/\omega_{pe}$, and $V_A=B/\sqrt{\mu_0 n_i m_i}$ is the ion Alfven velocity. The inifinite speed of light assumption further reduces the above condition  to $L_{0}\gg\left(r_{Le} \text{ and/or } \lambda_{e} \right)$, since $\lambda_{e}/\lambda_{De}=c/V_{Te}\gg1$.

After replacing the electric field, $\textbf{E}$, in the momentum equation (\ref{eq:2}) for ions using equation (\ref{eq:13}) and then applying the quasi-neutrality condition, a modified expression for the ion momentum equation can be written as:
\begin{equation}\label{eq:14}
\pD{\left(\rho_{i}\textbf{u}_{i}\right)}{t}+\nabla\cdot\left(\rho_{i}\textbf{u}_{i}\textbf{u}_{i}\right)=-\nabla{\left(p_{i}+p_{e}\right)}-\nabla\cdot\left(\mathbf{\pi}_{i}+\mathbf{\pi}_{e}\right)+\left(\textbf{R}_{e}+\textbf{R}_{i}\right)+\textbf{J}\times\textbf{B}+\rho_{i}\textbf{g}.
\end{equation}

\subsection{Final two-fluid plasma equations}
\label{sec:finaleq}

Finally, the two-fluid plasma transport equations considered in this study are the dimensional ion continuity equation, ion momentum equation, ion and electron internal energy equations, and Faraday’s law, and are summarized below as:
\begin{eqnarray}
\label{eq:15}
\pD{\rho_{i}}{t}+\nabla\cdot\left(\rho_{i}\textbf{u}_{i}\right)&=&0,\\
\label{eq:16}
\pD{\left(\rho_{i}\textbf{u}_{i}\right)}{t}+\nabla\cdot\left(\rho_{i}\textbf{u}_{i}\textbf{u}_{i}\right)&=&-\nabla{\left(p_{i}+p_{e}\right)}-\nabla\cdot\left(\mathbf{\pi}_{i}+\mathbf{\pi}_{e}\right)+\textbf{J}\times\textbf{B}+\rho_{i}\textbf{g},\\
\label{eq:17}
\pD{\left(\rho_{i}e_{i}\right)}{t}+\nabla\cdot\left(\rho_{i}e_{i}\textbf{u}_{i}\right)&=&-p_{i}\nabla\cdot\textbf{u}_{i}-\nabla\cdot\textbf{q}_{i}+\mathbf{\pi}_{i}:\nabla\textbf{u}_{i}+Q_{\Delta},\\
\label{eq:18}
\pD{\left(\rho_{i}e_{e}\right)}{t}+\nabla\cdot\left(\rho_{i}e_{e}\textbf{u}_{e}\right)&=&\left(\frac{m_{i}}{Zm_{e}}\right)\left[ -p_{e}\nabla\cdot\textbf{u}_{e}-\nabla\cdot\textbf{q}_{e}+\mathbf{\pi}_{e}:\nabla\textbf{u}_{e}+\right. \protect \nonumber \\
&&\left. \left(\frac{m_{i}}{eZ\rho_{i}}\right)\textbf{R}_{e}\cdot\textbf{J}-Q_{\Delta}\right],\\
\label{eq:19}
\pD{\textbf{B}}{t}&=&-\nabla\times\textbf{E},
\end{eqnarray}

\noindent
where the currently density, $\textbf{J}$, electron velocity, $\textbf{u}_{e}$, and electric field, $\textbf{E}$, are calculated from formulations (\ref{eq:10}), (\ref{eq:12}) and (\ref{eq:13}), respectively. The ion/electron pressures, $p_{s}$, and temperatures, $T_{s}$, are related through the ideal gas EOS as described in section(\ref{sec:twofluid}).        

As a result of infinite speed of light and negligible electron inertia assumptions, the consistency of quasi-neutrality condition ($\rho_{c}\approx 0$) in the final two-fluid plasma equations must be checked numerically by examining the value of charge density, $\rho_{c}$, calculated from equation (\ref{eq:6}). In other words, the divergence of the electric field, $\textbf{E}$, calculated from the generalized Ohm’s law equation (\ref{eq:13}) must be sufficiently small to maintain the quasi-neutrality condition. The numerical results obtained for all test cases confirm the quasi-neutrality condition and two sample results are presented in Appendix \ref{sec:neutrality}.    

\section{Non-Dimensional Analysis}
\label{sec:3}

In order to assess the importance of Hall and Biermann battery effects, resistivity, viscous stress, and heat flux to plasma flows in different regimes, as well as characterize special limiting cases, in this section a non-dimensional analysis of the two-fluid plasma equations is provided. In order to compare different applications, the characteristic scales that can be varied in practical problems of interests including temperature, number density, characteristic length scale, and magnetic field strength are chosen as the primary reference quantities. 

\subsection{Non-dimensional two-fluid plasma equations}
\label{sec:TwoND}

We choose the characteristic number density, $n_{0}$, length scale, $L_{0}$, temperature, $T_{0}$, magnetic field strength, $B_{0}$, and external acceleration, $g_{0}$, as the primary reference quantities and use them to construct scales for other variables like ion mass density, $\rho_{0}=n_{0}m_{i}$, ion Alfven velocity, $V_{A}^{0}=B_{0}/\sqrt{\mu_{0}n_{0}m_{i}}$, plasma pressure, $p^{0}=n_{0}k_{B}T_{0}$, a time scale, $L_{0}/V_{A}^{0}$, and so on. With these choices, the non-dimensional two-fluid plasma equations become:
\begin{eqnarray}
\label{eq:20}
\pD{\rho_{i}^{*}}{t^{*}}+\nabla^{*}\cdot\left(\rho_{i}^{*}\textbf{u}_{i}^{*}\right)&=&0,\\
\label{eq:21}
\pD{\left(\rho_{i}^{*}\textbf{u}_{i}^{*}\right)}{t^{*}}+\nabla^{*}\cdot\left(\rho_{i}^{*}\textbf{u}_{i}^{*}\textbf{u}_{i}^{*}\right)&=&-\beta\nabla^{*}\left(p_{i}^{*}+p_{e}^{*}\right)+\textbf{J}^{*}\times\textbf{B}^{*}-\frac{1}{Re^{i}}\nabla^{*}\cdot\mathbf{\pi}_{i}^{*}- \protect \nonumber  \\ 
&&\frac{1}{Re^{e}}\nabla^{*}\cdot\mathbf{\pi}_{e}^{*}+\frac{1}{Fr^{2}}\rho_{i}^{*}\textbf{g}^{*},\\                            
\label{eq:22}
\pD{\left(\rho_{i}^{*}e_{i}^{*}\right)}{t^{*}}+\nabla^{*}\cdot\left(\rho_{i}^{*}e_{i}^{*}\textbf{u}_{i}^{*}\right)&=&-\beta p_{i}^{*}\nabla^{*}\cdot\textbf{u}_{i}^{*}-\frac{\beta}{Re^{i}}\nabla^{*}\cdot\textbf{q}_{i}^{*}+\frac{1}{Re^{i}}\mathbf{\pi}_{i}^{*}:\nabla^{*}\textbf{u}_{i}^{*}+\protect \nonumber \\
&& 3Z\left(\frac{m_{i}}{m_{e}}\right)\beta\omega_{ei}^{0}Q_{\Delta}^{*},\\
\label{eq:23}
\pD{\left(\rho_{i}^{*}e_{e}^{*}\right)}{t^{*}}+\nabla^{*}\cdot\left(\rho_{i}^{*}e_{e}^{*}\textbf{u}_{e}^{*}\right)&=& \left(\frac{m_{i}}{Zm_{e}}\right)\left[-\beta p_{e}^{*}\nabla^{*}\cdot\textbf{u}_{e}^{*}-\left(\frac{m_{i}}{m_{e}}\right)\frac{\beta}{Re^{e}}\nabla^{*}\cdot\textbf{q}_{Te}^{*}-\right. \protect \nonumber \\
&&\left. \frac{1}{Re^{e}}\mathbf{\pi}_{e}^{*}:\nabla^{*}\textbf{u}_{e}^{*}+\frac{1}{Re_{m}}\textbf{R}_{\text{u}}^{*}\cdot\textbf{J}^{*}\right] -3\omega_{ei}^{0}\beta Q_{\Delta}^{*}+\protect \nonumber \\
&&\left(\frac{m_{i}}{m_{e}}\right)\hat{\lambda}_{i}\beta\left(-\nabla^{*}\cdot\textbf{q}_{ue}^{*}+\textbf{R}_{\text{T}}^{*}\cdot\textbf{J}^{*} \right),\\
\label{eq:24}
\pD{\textbf{B}^{*}}{t^{*}}&=&-\nabla^{*}\times\textbf{E}^{*},\\
\label{eq:25}
\textbf{E}^{*}&=&\hat{\lambda}_{i}\frac{1}{\rho_{i}^{*}}\left[\textbf{J}^{*}\times\textbf{B}^{*}-\beta\nabla^{*}p_{e}^{*}-\frac{1}{Re^{e}}\nabla^{*}\cdot\mathbf{\pi}_{e}^{*}+Z\beta\rho_{i}^{*}\textbf{R}_{\text{T}}^{*}+\right. \protect \nonumber \\
&&\left.\left(\frac{Zm_{e}}{m_{i}}\right)\frac{1}{Fr^{2}}\rho_{i}^{*}\textbf{g}^{*}\right]+ \frac{1}{Re_{m}}\textbf{R}_{\text{u}}^{*}-\textbf{u}_{i}^{*}\times\textbf{B}^{*},\\
\label{eq:26}
\textbf{u}_{e}^{*}&=&\textbf{u}_{i}^{*}-\hat{\lambda}_{i}\frac{1}{\rho_{i}^{*}}\textbf{J}^{*},\\
\label{eq:27}
\textbf{J}^{*}&=&\nabla^{*}\times\textbf{B}^{*}.
\end{eqnarray}

\noindent
where the superposed asterisk refers to the dimensionless variable and the non-dimensional parameters are the ion inertial scale or skin depth, 

\begin{equation}
\hat{\lambda}_{i}=c/\left(\omega_{pi}^{0}L_{0}\right)=m_{i}/\left(ZeL_{0}\sqrt{\mu_{0}\rho_{0}}\right), 
\end{equation}
ion and electron reference Reynolds (or viscous Lundquist) numbers, 
\begin{eqnarray}
Re^{i}&=&\rho_{0}V_{A}^{0}L_{0}/\mu_{i}^{0}, \\
Re^{e}&=&\left(\mu_{i}^{0}/\mu_{e}^{0}\right)Re^{i}, 
\end{eqnarray}
plasma \textit{beta}, 
\begin{equation}
\beta=n_{0}k_{B}T_{0}/\left(B_{0}^{2}/\mu_{0}\right),
\end{equation}
magnetic Reynolds (or resistive Lundquist) number, 
\begin{equation}
Re_{m}=\mu_{0}V_{A}^{0}L_{0}/\eta^{0},
\end{equation}
Froude number, 
\begin{equation}
Fr=V_{A}^{0}/\sqrt{g_{0}L_{0}},
\end{equation}
the collision frequency, 
\begin{equation}
\omega_{ei}^{0}=L_{0}/\left(V_{A}^{0}\tau_{e}^{0}\right).
\end{equation}
In the above non-dimensional parameters, $\eta^{0}=\left( \frac{m_{i}}{eZ\rho_{0}}\right)\times\left(\frac{m_{e}}{e\tau_{e}^{0}}\right)$ is the background resistivity, and the formulations for other reference variables are ion plasma frequency, $\omega_{pi}^{0}=Ze\sqrt{n_{0}/\left(\varepsilon_{0}m_{i} \right)}$, ion viscosity, $\mu_{i}^{0}=n_{0}k_{B}T_{0}\tau_{i}^{0}$, electron viscosity, $\mu_{e}^{0}=Zn_{0}k_{B}T_{0}\tau_{e}^{0}$, ion collision time, $\tau_{i}^{0}=\left[12\pi^{3/2}\varepsilon_{0}^{2}\sqrt{m_{i}}\left(k_{B}T_{0}\right)^{3/2}/\left(ln\Lambda e^{4}Z^4n_{0}\right)\right]$, and electron collision time, $\tau_{e}^{0}=\left[6\sqrt{2}\pi^{3/2}\varepsilon_{0}^{2}\sqrt{m_{e}}\left(k_{B}T_{0}\right)^{3/2}/\left(ln\Lambda e^{4}Z^2n_{0}\right)\right]$. The Coulomb logarithm ($ln\Lambda$) variation is described in Appendix \ref{sec:A}.
  
In addition, by using the relations, $r_{Le}/r_{Li}=\lambda_{e}/\lambda_{i}=\sqrt{m_{e}/m_{i}}$, we can summarize the range of applicability for the two assumptions made in this study in term of the non-dimensional ion length scales as: $\hat{\lambda}_{i}\ll\sqrt{m_{i}/m_{e}}$ and/or $\hat{r}_{Li}\ll\sqrt{m_{i}/m_{e}}$ depending on the local magnetic field strength. Thus, in magnetic dominant regime (e.g. low plasma $\beta$), the fact that $\hat{r}_{Li}/\hat{\lambda}_{i}=\sqrt{\beta}<1$ yields  $\hat{\lambda}_{i}\ll\sqrt{m_{i}/m_{e}}$. On the other hand, in plasma dominant regime (e.g. large plasma $\beta$), the applicability condition becomes $\hat{r}_{Li}\ll\sqrt{m_{i}/m_{e}}$.         

\subsection{Single-fluid limiting equations}
\label{sec:OneND}

In order to demonstrate the limiting cases of the two-fluid plasma equations solved in this study, the non-dimensional single-fluid plasma equations for ion-electron mixture density $\rho^{*}=\rho_{i}^{*}+\rho_{e}^{*}$, velocity, $\textbf{u}^{*}=\left(\rho_{i}^{*}\textbf{u}_{i}^{*}+\rho_{e}^{*}\textbf{u}_{e}^{*}\right)/\rho^{*}$, and pressure, $p^{*}=p_{i}^{*}+p_{e}^{*}$, are derived from the two-fluid plasma equations and given below:
\begin{eqnarray}
\label{eq:28}
\pD{\rho^{*}}{t^{*}}+\nabla^{*}\cdot\left(\rho^{*}\textbf{u}^{*}\right)&=&0,\\
\label{eq:29}
\pD{\left(\rho^{*}\textbf{u}^{*}\right)}{t^{*}}+\nabla^{*}\cdot\left(\rho^{*}\textbf{u}^{*}\textbf{u}^{*}\right)&=&-\beta\nabla^{*}p^{*}+\textbf{J}^{*}\times\textbf{B}^{*}-\frac{1}{Re^{i}}\nabla^{*}\cdot\mathbf{\pi}_{i}^{*}-\protect \nonumber \\
&&\frac{1}{Re^{e}}\nabla^{*}\cdot\mathbf{\pi}_{e}^{*}+\frac{1}{Fr^{2}}\rho^{*}\textbf{g}^{*}, \\                                
\label{eq:30}
\frac{1}{\gamma-1}\left[\pD{p^{*}}{t^{*}}+\nabla^{*}\cdot\left(p^{*}\textbf{u}^{*}\right)   \right]&=&-p^{*}\nabla^{*}\cdot\textbf{u}^{*}+\frac{\hat{\lambda}_{i}}{\gamma-1}\nabla^{*}\cdot\left(\frac{p_{e}^{*}}{\rho^{*}}\textbf{J}^{*}\right)+\hat{\lambda}_{i}p_{e}^{*}\nabla^{*}\cdot\left(\frac{\textbf{J}^{*}}{\rho^{*}}\right)+ \protect \nonumber \\ &&\frac{1}{\beta Re_{m}}\textbf{R}_{\text{u}}^{*}\cdot\textbf{J}^{*}- \left(\frac{1}{Re^{i}}\nabla^{*}\cdot\textbf{q}_{i}^{*}+ \frac{m_{i}}{m_{e}}\frac{1}{Re^{e}}\nabla^{*}\cdot\textbf{q}_{Te}^{*}\right)-\protect \nonumber \\
&&\frac{1}{\beta}\left(\frac{1}{Re^{i}}\mathbf{\pi}_{i}^{*}:\nabla^{*}\textbf{u}_{i}^{*}+\frac{1}{Re^{e}}\mathbf{\pi}_{e}^{*}:\nabla^{*}\textbf{u}_{e}^{*}\right)+ \protect \nonumber \\
&&Z\hat{\lambda}_{i}\left(-\nabla^{*}\cdot\textbf{q}_{ue}^{*}+\textbf{R}_{\text{T}}^{*}\cdot\textbf{J}^{*}\right).                                                   
\end{eqnarray}

\noindent
Equation (\ref{eq:28}) is obtained by applying the relations $\rho_{e}^{*}/\rho_{i}^{*}=Zm_e/m_i$, $\rho^{*}=\rho_{i}^{*}+\rho_{e}^{*}$, $\textbf{u}^{*}=\left(\rho_{i}^{*}\textbf{u}_{i}^{*}+\rho_{e}^{*}\textbf{u}_{e}^{*}\right)/\rho^{*}$, and equations (\ref{eq:26}) and (\ref{eq:27}) into equation (\ref{eq:20}). Similarly, by using the above  ion-electron mixture variables definitions (including $p^{*}=p_{i}^{*}+p_{e}^{*}$) and equation (\ref{eq:26}), one can obtain equation (\ref{eq:29}) from the ion momentum equation (\ref{eq:21}) under the negligible electron inertia assumption. Finally, using the non-dimensional EOS, $\rho_{s}^{*}e_{s}^{*}=\beta/\left(\gamma-1\right)p_{s}^{*}$, the ion-electron mixture variables definitions, negligible electron inertia assumption, and equation (\ref{eq:26}), equation (\ref{eq:30}) is obtained by taking summation of ion and electron energy equations (\ref{eq:22}) and (\ref{eq:23}).     

In addition, the generalized Ohm’s law is rewritten as:
\begin{eqnarray}\label{eq:31}
\textbf{E}^{*}+\textbf{u}^{*}\times\textbf{B}^{*}&=&\hat{\lambda}_{i}\frac{1}{\rho^{*}}\textbf{J}^{*}\times\textbf{B}^{*}-\hat{\lambda}_{i}\frac{\beta}{\rho^{*}}\nabla^{*}p_{e}^{*}+\frac{1}{Re_{m}}\textbf{R}_{\text{u}}^{*}-\hat{\lambda}_{i}\frac{1}{\rho^{*}Re^{e}}\nabla^{*}\cdot\mathbf{\pi}_{e}^{*}+\protect \nonumber \\
&&Z\hat{\lambda}_{i}\beta\textbf{R}_{\text{T}}^{*}+\hat{\lambda}_{i}\frac{Z}{Fr^{2}}\frac{m_{e}}{m_{i}}\textbf{g}^{*}.
\end{eqnarray}

\noindent
Equation (\ref{eq:31}) is obtained by replacing the ion-electron mixture variables and equation (\ref{eq:26}) into equation (\ref{eq:25}). The Faraday’s law for the non-dimensional magnetic field, $\textbf{B}^{*}$, and the reduced Ampere’s law for current density, $\textbf{J}^{*}$, remain unchanged as equations (\ref{eq:24}) and (\ref{eq:27}), respectively. 

The ion velocity $\textbf{u}_{i}^{*}$ and electron velocity $\textbf{u}_{e}^{*}$ can be obtained by using the relations  $\rho_{e}^{*}/\rho_{i}^{*}=Zm_e/m_i$, $\rho^{*}=\rho_{i}^{*}+\rho_{e}^{*}$, $\textbf{u}^{*}=\left(\rho_{i}^{*}\textbf{u}_{i}^{*}+\rho_{e}^{*}\textbf{u}_{e}^{*}\right)/\rho^{*}$, and the definition of current density under quasi-neutrality condition (\ref{eq:26}) and then are used to calculate the viscous stresses, $\mathbf{\pi}_{e}^{*}$ and $\mathbf{\pi}_{i}^{*}$, appearing in the single-fluid equations (\ref{eq:29}), (\ref{eq:30}) and (\ref{eq:31}). 

Written as above, the equations are unclosed, as the ion and electron temperatures and densities can not be independently determined. For plasma flows with identical ion and electron temperatures (i.e. $T_{i}^{*}= T_{e}^{*}$ ), the ion and electron temperatures become the same as the mixture temperature, $T^{*}$, which can be obtained from the total pressure, $p^*$, by using the EOS. In addition, the electron pressure can be obtained via the relation $p_{e}^{*}=Z p_{i}^{*}=p^*/\left(1+1/Z\right)$. In this case, the single-fluid plasma equations, including all transport terms, are closed. 

As demonstrated in Appendix C, the conventional Hall, resistive, and ideal MHD equations can be recovered from the above single-fluid plasma equations,  as limiting cases, in regimes where the non-dimensional parameters satisfy the corresponding conditions described below:

\begin{itemize}
\item The conventional Hall-MHD equations can be recovered in regimes where $Re^{i},\ Re^{e}\rightarrow\infty$, $Fr\rightarrow\infty$, $\textbf{q}_{ue}^{*} \rightarrow\ 0$, and $\textbf{R}_{T}^{*} \rightarrow\ 0$ ($\textbf{q}_{ue}^{*}$ and $\textbf{R}_{T}^{*}$ are always ignored in the Hall-MHD equations and are only considered in Braginskii two-fluid model~\cite{braginskii1965}). 
\item Resistive MHD equations can be recovered in regimes where $\hat{\lambda}_{i}\rightarrow0$ (and/or $\hat{r}_{Li}\rightarrow0$), $Re^{i},\ Re^{e}\rightarrow\infty$, and $Fr\rightarrow\infty$. 
\item Ideal MHD equations can be recovered in regimes where $\hat{\lambda}_{i}\rightarrow0$ (and/or $\hat{r}_{Li}\rightarrow0$), $Re^{i},\ Re^{e}\rightarrow\infty$, $Fr\rightarrow\infty$, and $Re_{m}\rightarrow\infty$.
 \end{itemize}

\noindent
The Hall-MHD equations can sometimes be classified as a two-fluid model because of the inclusion of the Hall term and electron pressure gradient. However, similar like the discussion made above, due to the presence of the electron pressure, the Hall-MHD equations are not closed. In practice, to close the Hall-MHD equations, some studies \cite{Callen2006,Dmitruk2006,Toth2008} simply neglect the electron pressure, while others \cite{Browning2014,Srinivasan2011} assume identical ion and electron temperatures ($T_{i}^{*}= T_{e}^{*}$) and obtain the electron pressure as $p_{e}^{*}=Z p_{i}^{*}=p^*/\left(1+1/Z\right)$. 

The viscous terms vanish from the single-fluid equations in the limit of infinite Reynolds numbers only for non-turbulent flows. This restricts the domain of applicability of the limiting cases above, unless models for subgrid or turbulence transport are added to the equations. In some recent studies~\cite{Dmitruk2006,Browning2014}, viscous effects are included in the Hall-MHD equations for regimes where $Re^{i}$ is not sufficiently large. However, the formulations for the viscous terms are more or less ad-hoc. Some studies~\cite{Dmitruk2006}, model the viscous stress term by using ion-electron mixture velocity, $\textbf{u}^{*}$, with standard formulation for compressible ideal gas, i.e. $\nabla^{*}\cdot\mathbf{\pi}^{*}=\nabla^{*2}\textbf{u}^{*}+(1/3)\nabla^{*}\nabla^{*}\cdot\textbf{u}^{*}$ instead of the detailed plasma formulations for $\mathbf{\pi}_{i}^{*}$ and $\mathbf{\pi}_{e}^{*}$ given in Appendix \ref{sec:A}. In addition, the viscous contribution was only added to the momentum equations.

In the next section, numerical simulations will be conducted for a series of canonical problems to highlight the accuracy and robustness of the two-fluid plasma solver in handling plasma flows in different regimes. 

\section{Test cases}
\label{sec:4}

The dimensional two-fluid plasma equation with full transport terms described in section \ref{sec:finaleq} have been implemented in the petascale CFDNS code~\cite{cfdns,petersen2010fss,ryu2014} and solved by using sixth-order non-dissipative compact finite differences~\cite{Lele92,petersen2010fss} for four canonical problems: Alfven and whistler dispersion relations, electromagnetic plasma shock, and magnetic reconnection. For these cases, ion and electron temperatures are the same, i.e. $T_{i}=T_{e}$. Therefore, the collision generated heat for ion energy equation, $Q_{i}\left(Q_{\Delta}\right)$, vanishes while the collision generated heat for electron energy equation, $Q_{e}$, reduces to the ohmic heating term shown as the fourth term in the RHS of equation (\ref{eq:18}). Therefore, the two-fluid plasma equations solved in these test cases are mathematically equivalent to the single-fluid plasma equations described in section \ref{sec:OneND} which can be viewed as the general or full Hall-MHD equations (therefore more general than the conventional Hall-MHD equations used in previous studies and explained in Appendix \ref{sec:Hall-MHD}) including all plasma transport terms. The identical temperature simplification further eliminates the need to solve the ion energy equation (\ref{eq:17}). 

For the test cases considered in this study, the initial conditions for all primary variables (non-dimensional) are identical to those given in the references mentioned below. The values of the non-dimensional parameters, i.e., $\hat{\lambda}_{i}$, $Re^{i}$ and $Re_{m}$, are calculated based on the characteristic number density, $n_{0}$, length scale, $L_{0}$, temperature, $T_{0}$, and magnetic strength, $B_{0}$, and chosen to match previous studies and/or certain practical applications, with the requirement that the simulations remain well-resolved.

\subsection{Alfven and whistler dispersion relations}
\label{sec:Wave}

The first two test cases used to test the accuracy of the newly developed two-fluid plasma solver are the dispersion relations for Alfven and whistler waves. These are two plasma phenomena often observed in different space flow regimes~\cite{Priest1984,Tomczyk2007,Stenzel1999,Cassak2012}. The frequency and length scales for Alfven waves satisfy the relations $\omega\ll\omega_{ci}$ and $L_{0}\gg c/\omega_{pi}=\lambda_{i}$~\cite{Bellan2006}. Therefore, Alfven waves become ideal MHD waves when the local $Re^{i}$ and $Re_{m}$ values are sufficiently high. The basic frequency and length scales for whistler waves are in the ranges of $\omega_{ci}\ll\omega\ll\omega_{ce}$ and $\lambda_{e}\ll L_{0}\ll\lambda_{i}$~\cite{Stenzel1999,Bellan2006}. Therefore, whistler waves are a Hall-MHD phenomenon.  

By linearizing the ideal MHD and Hall-MHD equations about the equilibrium and assuming plane wave solutions of the form $exp\left(ik^{*}x^{*}-i\omega^{*}t^{*}\right)$, one obtains the Alfven and whistler dispersion relations: 
\begin{eqnarray}\label{eq:32}
\omega^{*}&=&k^{*}\quad \text{for Alfven waves,}\\
\label{eq:33}
\left(\omega^{*2}-k^{*2}\right)^{2}&=&\omega^{*2}k^{*4}\quad \text{for Whistler waves,}
\end{eqnarray}

\noindent
where $k^{*}=2\pi m/L_{x}^{*}$ is the wavenumber, $m$ is the integer mode, and $\omega^{*}$ is the wave frequency. The initial conditions are:
\begin{align*}
\rho_{i}^{*}=1,\qquad & u_{i}^{*}=0,\qquad & v_{i}^{*}=-\delta^{*}\cos\left(k^{*}x^{*}\right),\qquad & w_{i}^{*}=\delta^{*}\sin\left(k^{*}x^{*}\right),\qquad
\\p_{i}^{*}=1,\qquad & B_{x}^{*}=1,\qquad & B_{y}^{*}=\delta^{*}v_{p}^{*}\cos\left(k^{*}x^{*}\right),\qquad & B_{z}^{*}=-\delta^{*}v_{p}^{*}\sin\left(k^{*}x^{*}\right),\qquad
\end{align*}

\noindent
where $v_{p}^{*}$ is the phase velocity which can be calculated from linear equations (\ref{eq:32}) and (\ref{eq:33}). The simulations are conducted over a periodic domain with size $L_{x}^{*}=9.6$ and number of grid points $NX=384$. Therefore, for the simulations conducted in this study, the largest wave resolution is 192 points per wavelength for the minimum mode (i.e. $m=2$). For the case using the maximum mode (i.e. $m=30$), the wave resolution becomes $12.8$ ($=384/30$) points per wavelength. The initial perturbation amplitude $\delta^{*}=\text{10}^{\text{-5}}$ was used in all simulations.  

\begin{figure}[h!t]
\includegraphics[trim=70 512 70 75, clip, width=6.4in]{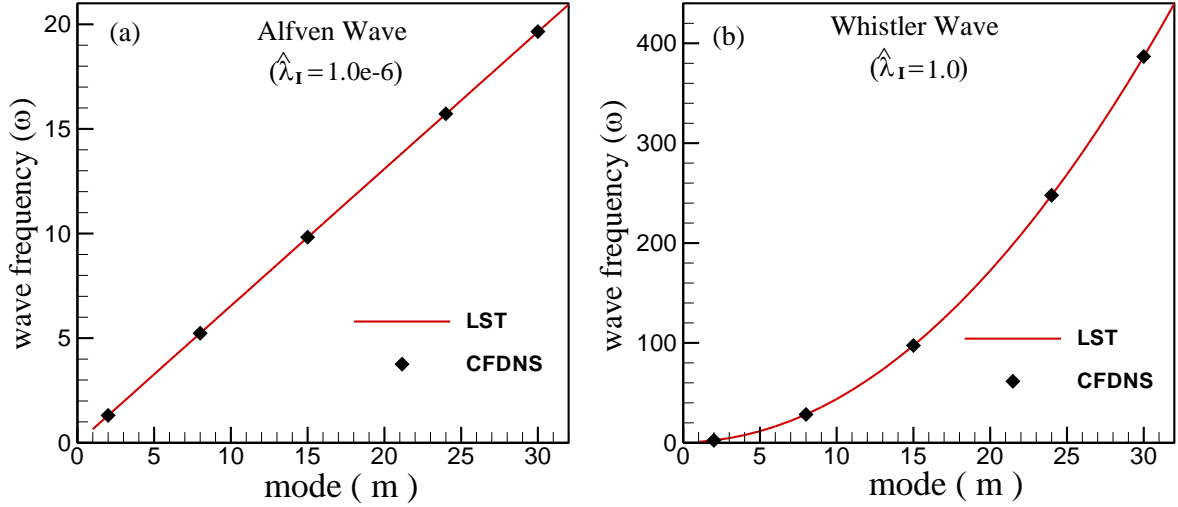}
\caption{Comparison of analytical Alfven and whistler dispersion relations (LST) with numerical solutions calculated by using the two-fluid plasma solver (CFDNS) in (a) ideal MHD regime and (b)Hall-MHD regime.}
\label{fig:Wave}
\end{figure}

Because Alfven waves are believed to be the main mechanism for heating the solar corona~\cite{Priest1984}, the characteristic number density, $n_{0}$, length scale, $L_{0}$, temperature, $T_{0}$, and magnetic strength, $B_{0}$, are chosen as the typical values for solar corona/flares~\cite{Parker1983,Browning2014}. These characteristic values are $L_{0}\sim\text{10}^7\text{m}$, $n_{0}\sim\text{10}^{15}\text{m}^{-3}$, $T_{0}=100\text{eV}$ and $B_{0}\sim0.01 \text{T}$, which leads to $\hat{\lambda}_{i}\sim 1.0\times{10}^{-6}$, $Re_{m}\sim 2.0\times{10}^{13}$, and $Re^{i}\sim 1.0\times{10}^{4}$. Fig.~\ref{fig:Wave}$(a)$ shows that the CFDNS results calculated using the two-fluid plasma equations in the global solar corona regime are in excellent agreement with the analytical linear stability theory (LST) predictions for ideal MHD over a wide range of modes, m.  

The whistler waves found in solar corona/flares are related to the fast, collisionless magnetic reconnection that occurs on the length-scales comparable with the ion skin-depth~\cite{Mandt1994,Cassak2012}. The ion-skin depth estimated using the typical parameter values of solar corona/flare parameters shown above is similar to the one provided in Ref.~\cite{Browning2014}, i.e. $\lambda_{i}\approx 10\text{m}$.  However, the viscous effects estimated using the closures described in Ref.~\cite{braginskii1965} and Appendix \ref{sec:A} are probably not accurate at this scale, which is smaller than the mean free path for the solar corona/flares~\cite{Parker1983,Browning2014}. Developing closures applicable to collisionless systems is difficult~\cite{Wang2015}. Therefore, to be able to perform simulations relevant to the whistler wave dispersion relation, yet maintain the correspondence to the solar corona/flares parameters, we still use the above parameters, but decrease the reference temperature $T_0$ to obtain high enough values of $Re^{i}\approx 4.0\times10^3$ and $Re_{m}\approx 1.0\times10^4$. Again, as shown in Fig.~\ref{fig:Wave}$(b)$, the CFDNS results calculated using the two-fluid plasma equations perfectly match the analytical solution given by Eq. (\ref{eq:33}). 

\subsection{Electromagnetic plasma shock}
\label{sec:Shock}

The presence of plasma shocks is also often observed in space and fusion applications. For example, the interaction of the solar wind with the Earth magnetosphere leads to the formation of a bow shock upstream of the magnetopause~\cite{Shen1972,Peredo1995}. The electromagnetic plasma shock simulated here is an extension of the single-fluid, inviscid Brio-Wu shock~\cite{Brio1988} to the  two-fluid plasma model. The initial values for the non-dimensional primary variables are: 
\begin{equation*}
\begin{bmatrix}
  \rho_{i}^{*}\\
  u_{i}^{*}\\
  v_{i}^{*}\\ 
  w_{i}^{*}\\
  p_{i}^{*}\\
  B_{x}^{*}\\
  B_{y}^{*}\\
  B_{z}^{*}\\
\end{bmatrix}
=
\begin{bmatrix}
  1.0\\
  0.0\\
  0.0\\
  0.0\\
  0.5\\
  0.75\\
  1.0\\
  0.0\\
\end{bmatrix}
\: \text{For} \: x^{*}\leq0.5 \qquad \text{and} \qquad
\begin{bmatrix}
  \rho_{i}^{*}\\
  u_{i}^{*}\\
  v_{i}^{*}\\ 
  w_{i}^{*}\\
  p_{i}^{*}\\
  B_{x}^{*}\\
  B_{y}^{*}\\
  B_{z}^{*}\\
\end{bmatrix}
=
\begin{bmatrix}
  0.125\\
  0.0\\
  0.0\\
  0.0\\
  0.05\\
  0.75\\
  -1.0\\
  0.0\\
\end{bmatrix}
\: \text{For} \: x^{*}> 0.5
\end{equation*}

\noindent
and Dirichlet boundary conditions are implemented at the shock-tube boundaries. 

\begin{figure}[h!t]
\includegraphics[trim=70 330 78 75, clip, width=6.0in]{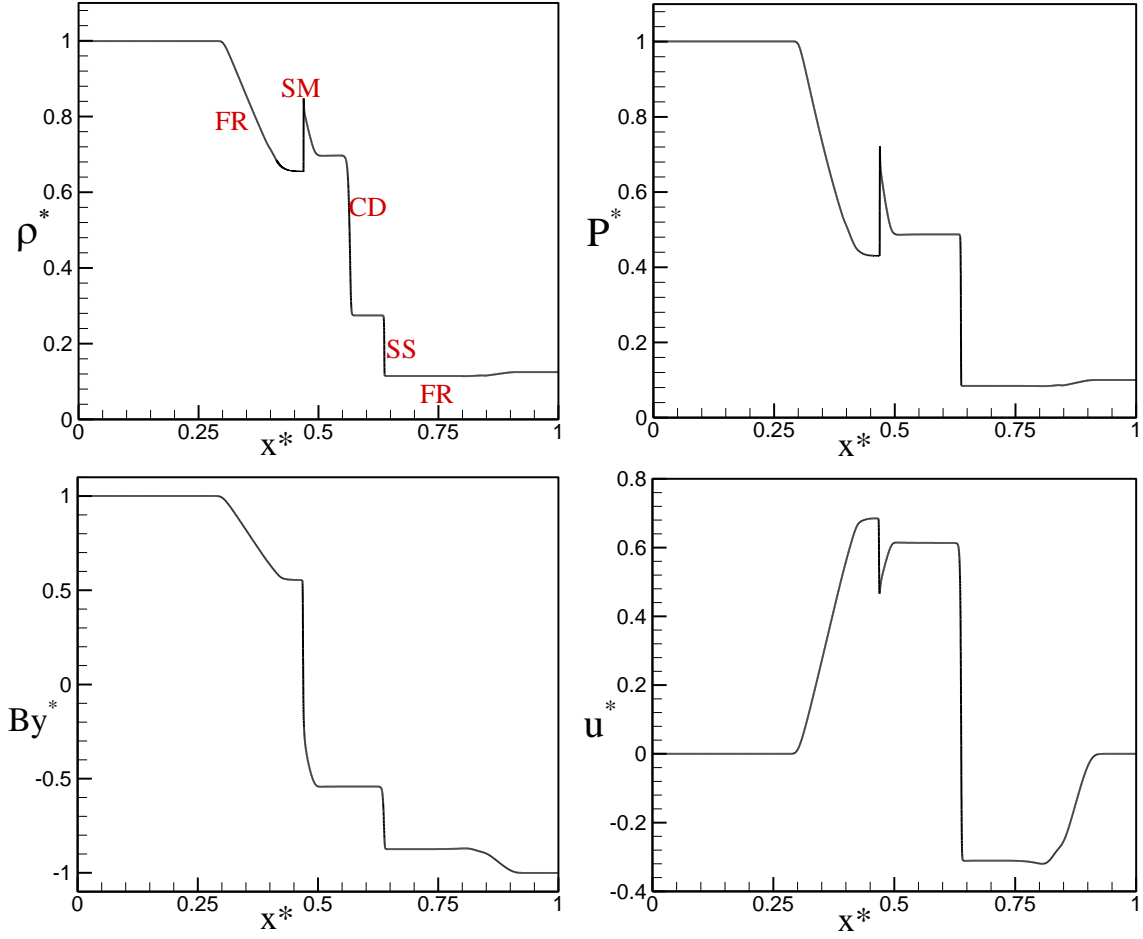}
\caption{CFDNS results calculated by using two-fluid plasma equations for plasma shock with normalized ion skin depth, $\hat{\lambda}_{i}=1.0\times10^{-6}$, magnetic Reynolds number, $Re_m\sim1.0\times10^{12}$, and ion viscous Reynolds number, $Re^{i}=6.3\times10^{3}$.}
\label{fig:shock1}
\end{figure} 

Most (if not all) previous numerical studies of the bow shock~\cite{Mejnertsen2018,Gombosi1994,Chapman2004} and Brio-Wu shock~\cite{Jiang1999,Brio1988,Loverich2011} ignore the viscous and heat flux terms and fully rely on the numerical dissipation introduced by shock-capturing schemes to regularize the equations around sharp discontinuities. On the contrary, by including full plasma transport terms, one should be able to resolve the shocks by using high-order non-dissipative numerical schemes at sufficiently high grid resolution. Therefore, in this test case, we choose the characteristic number density, $n_{0}\sim 10^7\text{m}^{-3}$, length scale, $L_{0}\sim 10^{11}\text{m}$, and magnetic strength, $B_{0}\sim 10\:\text{nT}$, as the typical values found in solar wind~\cite{Mejnertsen2018,Bellan2006,Chapman2004},  and vary $T_0$ to obtain a range of substantial high, but still affordable Reynolds numbers. These reference scales give $\hat{\lambda}_{i}\sim 1.0\times{10}^{-6}$ and $Re_{m}\sim 10^{12}$, therefore, both Hall effect and magnetic resistivity become negligible. 

Fig.~\ref{fig:shock1} shows that, without the need to explicitly turn off the corresponding terms from the governing equations, the ideal MHD results including the slow compound wave (SM), contact discontinuity (CD), and slow shock (SS) are obtained from the two-fluid plasma solver for flows with large enough Reynolds numbers. With the presence of physical viscosity, the shock wave is no longer zero-thickness. Instead, the value of shock thickness depends on the local Reynold number. Therefore, all shock structures can be fully resolved by using high-order non-dissipative numerical schemes provided the grid resolutions are sufficiently high. Of course, by increasing viscosity or decreasing the Reynolds number, the profiles for all variables become smoother and, therefore, can easily be resolved at lower grid resolutions. 

\begin{figure}[h!t]
\includegraphics[trim=74 525 81 77, clip, width=6.4in]{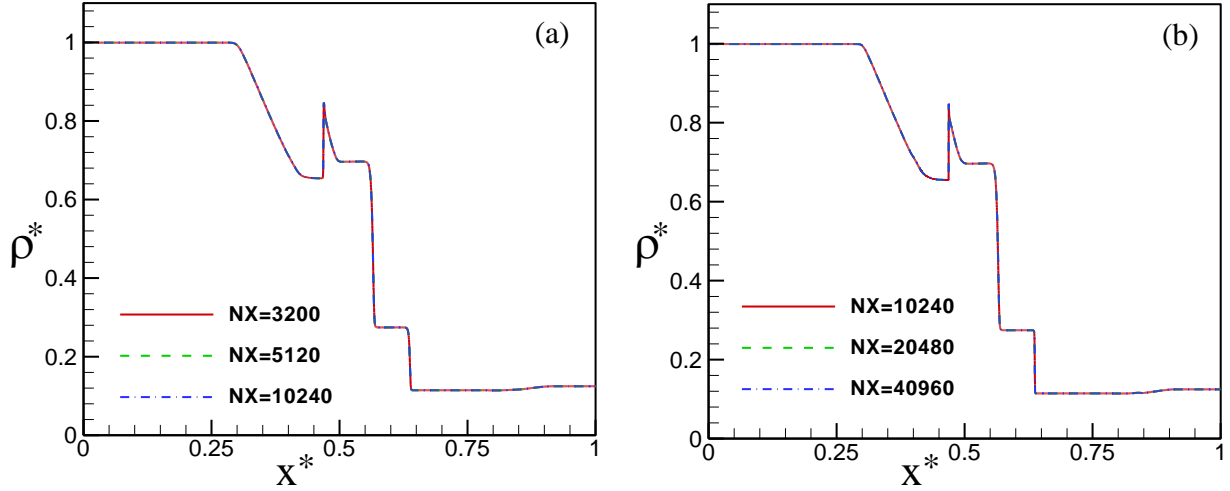}
\caption{Fully converged CFDNS results for plasma shocks at two ion viscous Reynolds numbers (a) $Re^{i}=6.3\times10^{3}$ and (b) $Re^{i}=2.3\times10^{4}$.}
\label{fig:shock2}
\end{figure} 

In this study, grid convergence tests have been conducted for all plasma shock cases to guarantee that computational results presented are free of numerical error.  As indicated in Fig.~\ref{fig:shock2}, fully resolved DNS-like solutions are obtained at all ion viscous Reynolds number when the ion grid Reynolds number, $Re_{\Delta}=Re^{i}/NX$, is smaller than a threshold value, which is $2.3$ for our 6th order compact finite differences solver. 

\begin{figure}[h!b]
\includegraphics[trim=80 520 73 72, clip, width=6.4in]{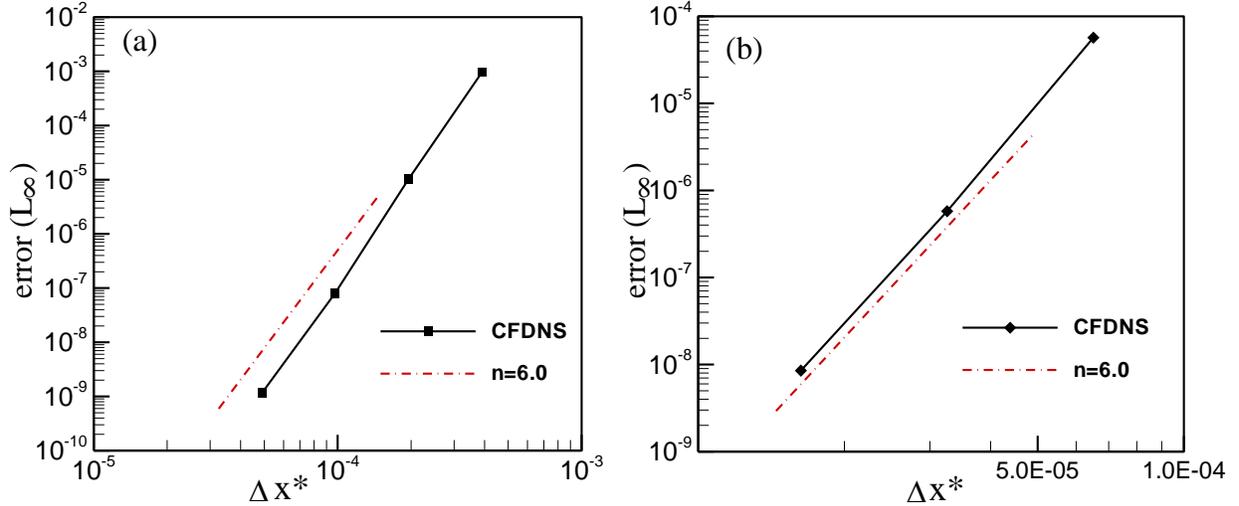}
\caption{The comparison of grid convergence rates calculated by two-fluid plasma solver for plasma shock at two ion viscous Reynolds numbers (a) $Re^{i}=6.3\times10^{3}$ and (b) $Re^{i}=2.3\times10^{4}$ with the theoretical limit of sixth-order compact scheme ($n=6$).}
\label{fig:shock3}
\end{figure} 

By using the finest grid solutions as the “exact” results, i.e. $NX=40960$ for $Re^{i}=6.3\times10^3$ and $NX=122880$ for $Re^{i}=2.3\times10^4$, Fig.~\ref{fig:shock3} shows that the grid convergence rates are $\hat{n}=6.14$ for $Re^{i}=6.3\times10^3$ case and $\hat{n}=6.08$ for $Re^{i}=2.3\times10^4$ cases. Both values are very close to the theoretical limit of sixth-order compact finite differences scheme. This shows that the high-order two-fluid plasma solver results are free of the spurious behavior commonly found in high-order shock-capturing schemes results, like the modification of discontinuity location~\cite{Yee2013}. Secondly, the CFDNS results maintain nearly 6-order accuracy across the discontinuities, while the convergence rate of most shock-capturing schemes drops to first-order accuracy near discontinuities~\cite{Greenough2004}. 

\subsection{Magnetic reconnection}
\label{sec:Reconnection}

The last test case considered in this study is the collisionless magnetic reconnection, a rapid rearrangement of magnetic field topology and release of free magnetic energy. It is of particular importance to the dynamic evolution of the solar corona/flares~\cite{Priest1984,Cassak2012}, the magnetosphere~\cite{Nagai2001,Deng2001}, and thermonuclear fusion~\cite{Wesson1990,Browning2016}. Previous studies~\cite{Mandt1994,Birn2001,Shay2001,Ma2001,Otto2001} confirm that the fast magnetic reconnection occurs on a length scale comparable to ion skin depth and is mainly contributed by the Hall term.

Though extensive computational work has been done on the magnetic reconnection problem, simulations of magnetic reconnection with explicit viscous and thermal diffusion effects are rare. In addition, instead of a dynamically changing property, the resistivity in most previous studies~\cite{Toth2008,Ma2001} was simply chosen as a constant value. The justification for the absence of physical plasma transport terms is partially because the rapid magnetic reconnection is collisionless, therefore, the closures for transport terms based on Chapman-Enskog expansion in small mean-free-path~\cite{braginskii1965} become inappropriate, while developing closures applicable to collisionless systems is difficult~\cite{Wang2015}. In turn, most widely used plasma solvers use dissipative shock-capturing techniques and rely on numerical dissipation instead of physical transport terms to regularize the equations. In general, in such approaches the numerical dissipation is related to the mesh size and the simulations do not converge as the mesh size is increased. Therefore, it seems impossible for such plasma flow solvers to produce fully resolved DNS-like solutions. 

\begin{figure}[h!b]
\includegraphics[trim=110 512 105 75, clip, width=5.0in]{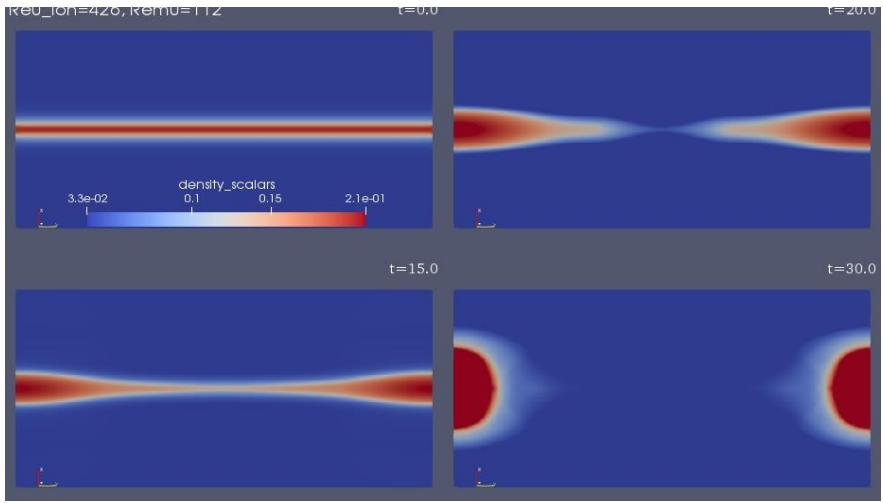}
\caption{The temporal variations of density contours during magnetic reconnection with $Re_{m}\approx 112$ and $Re^{i}\approx 426$.}
\label{fig:reconnection1}
\end{figure} 

In this study, we choose the characteristic number density and length scale as the typical values found in solar flare reconnection\cite{Browning2014}, i.e. $L_{0}\sim\text{10}\text{m}$ and $n_{0}\sim 10^{15}\text{m}^{-3}$, which leads to $\hat{\lambda}_{i}=1.0$. We demonstrate that physical transport can be used to obtain mesh converged solutions with negligible numerical dissipation. Moreover, as 
the viscous Reynolds number is increased, the solutions tend to converge and predict the colisionless  magnetic reconnection results. We vary the reference temperature, $T_0$, and magnetic strength, $B_0$, to generate a wide range of viscous Reynolds number, $Re^{i}$, and magnetic Reynolds number, $Re_{m}$, values. The range of $Re_{m}$ values is chosen to include values used in previous studies, i.e. $Re_{m}=100\sim350$  and $Re_{m}=200$ employed by Ma~\cite{Ma2001} and Toth~\cite{Toth2008}, respectively. 

\begin{figure}[h!t]
\includegraphics[trim=115 522 115 65, clip, width=5.0in]{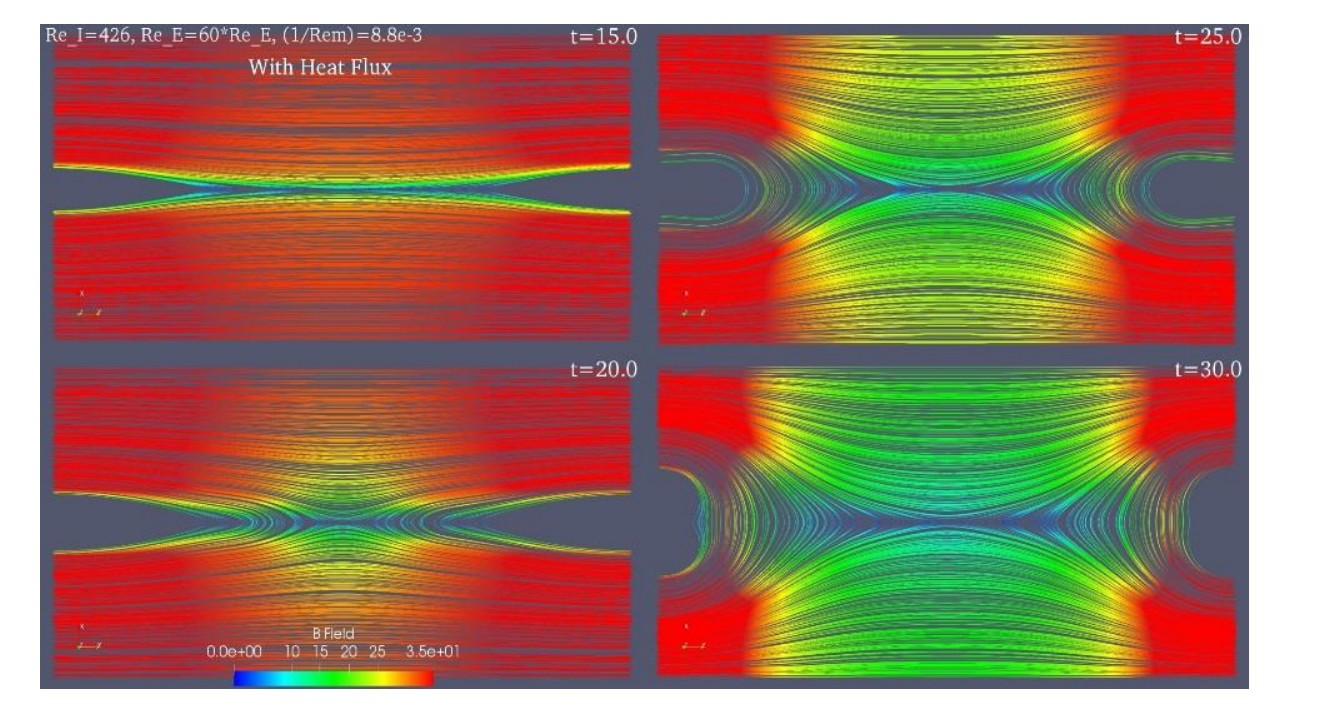}
\caption{The temporal variations of magnetic streamlines during magnetic reconnection with $Re_{m}\approx 112$ and $Re^{i}\approx 426$.}
\label{fig:reconnection2}
\end{figure} 

Similar to previous studies, the initial conditions for the non-dimensional primary variables are:  
\begin{equation*}   
\begin{bmatrix}
 \rho_{i}^{*}\\
  u_{i}^{*}\\
  v_{i}^{*}\\ 
  w_{i}^{*}\\
  p_{i}^{*}\\
  B_{x}^{*}\\
  B_{y}^{*}\\
  B_{z}^{*}\\
\end{bmatrix}
=
\begin{bmatrix}
  1/5+\text{sech}^2\left(2x^{*}\right)\\ 
  0.0\\
  0.0\\
  0.0\\
  \left[ 1/5+\text{sech}^2\left(2x^{*}\right)\right]/4 \\
  \left(2\pi/10L_{y}^{*}\right)\times\sin\left(2\pi y^{*}/L_{y}^{*}\right)\times\cos\left(\pi x^{*}/L_{x}^{*}\right) \\
  \tanh\left(2x^{*}\right)-\left(\pi/10L_{x}^{*}\right)\times\cos\left(2\pi y^{*}/L_{y}^{*}\right)\times\sin\left(\pi x^{*}/L_{x}^{*}\right) \\
  0.0\\
\end{bmatrix}
\end{equation*}

\noindent
The perfectly conducting wall boundary condition is applied in the vertical direction ($x^{*}$) and the periodic boundary condition is implemented in the horizontal direction ($y^{*}$). The simulations are conducted in a two-dimensional domain with $L_{x}^{*}=12.8$ and $L_{x}^{*}=25.6$.

Figs.~\ref{fig:reconnection1} and ~\ref{fig:reconnection2} show that the two-fluid non-dissipative plasma solver (i.e. CFDNS) with temperature and magnetic field dependent transport (ion/electron viscous stress, heat flux, frictional drag force, and magnetic resistivity) can successfully reveal the whole magnetic reconnection process. For example, Fig.~\ref{fig:reconnection1} indicates that, during the reconnection process, the high-density sheet is stretched and finally broken up into two ligaments, which further shrink to increase the high density values. In comparison to previous two-fluid plasma results without transport terms~\cite{Srinivasan2011}, the CFDNS results remain perfectly symmetric, which indicates the high accuracy of the two-fluid plasma solver in handling this challenging plasma flow. Fig.~\ref{fig:reconnection2} clearly shows that, as the reconnection takes place, the magnetic streamlines tend to bend from horizontal direction to vertical direction and the intensity of vertical component, $B_{x}$, increases dramatically. This corresponds to a rapid increase of reconnection flux and an eruptive release of magnetically stored energy to heat the plasma.

\begin{figure}[h!b]
\includegraphics[trim=115 512 142 65, clip, width=5.0in]{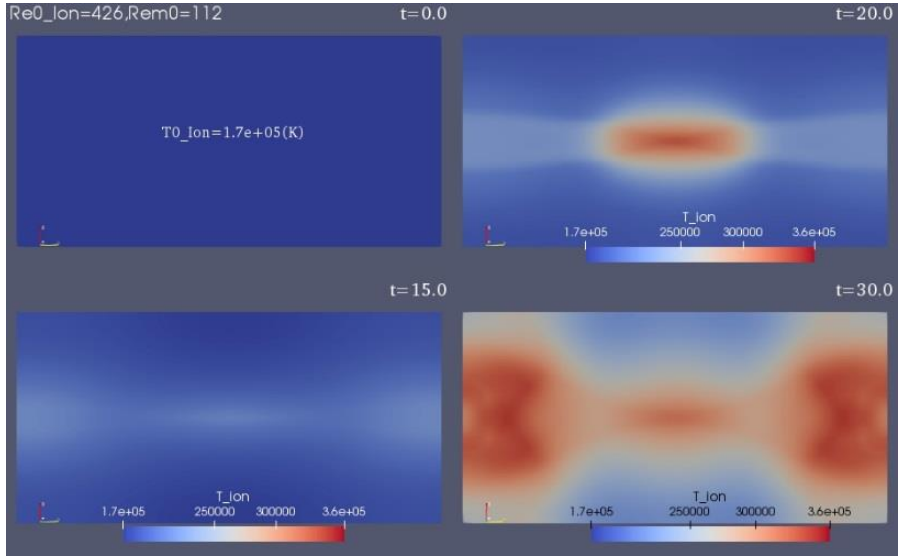}
\caption{The temporal variations of magnetic streamlines during magnetic reconnection with $Re_{m}\approx 112$ and $Re^{i}\approx 426$. }
\label{fig:reconnection3}
\end{figure} 

The temperature contours shown in Fig.~\ref{fig:reconnection3} further confirm the rapid conversion of magnetic energy into particle energy. As the reconnection takes place, both temperature and velocities (not shown) increase significantly due to the rapid conversion of magnetic energy into thermal and kinetic energies. In the solar corona, this phenomenon is thought to give rise to solar flares and drive the outflow of the solar wind~\cite{Priest1984}.  Consistently, the rapid increase of temperature causes a dramatic increase of heat flux and viscous dissipation, since $\kappa\propto T^{5/2}$ and $\mu^{s}\propto T^{5/2}$, as well as a large decrease of magnetic resistivity, since $\eta\propto T^{-3/2}$. 
The presence of thermal diffusion is then absolutely necessary to prevent unphysically high temperatures to be generated at the reconnection points. Previous studies without physical thermal diffusion had to rely on the numerical diffusion introduced by dissipative numerical schemes to damp this effect. The effect of numerical diffusion is hard quantify due to the higher order nonlinearities  usually present in the associated terms (if such can be explicitly evaluated at all). In addition, different numerical schemes have different truncation errors, so numerical diffusion is difficult to generalize across various codes. Thus, numerical results relying on numerical diffusion to regularize the equations should be regarded with caution.   

\begin{figure}[h!b]
\includegraphics[trim=75 510 70 70, clip, width=6.4in]{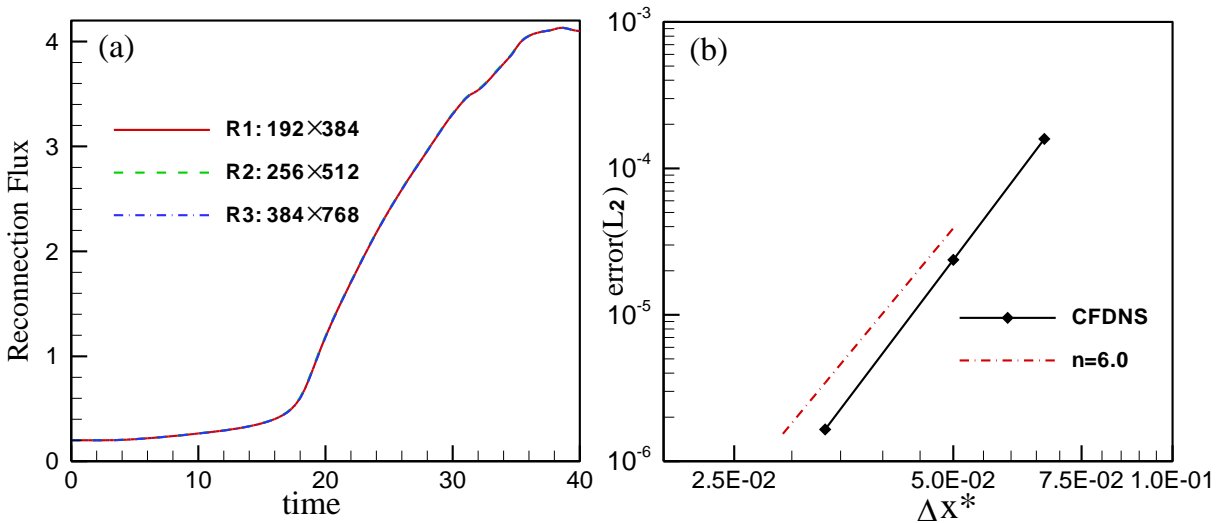}
\caption{(a) Temporal variation of reconnection flux at different resolutions and (b) comparison of grid convergence rate for the two-fluid plasma solver results ($\hat{n}=6.58$) against the theoretical limit of the sixth-order compact finite difference scheme. The simulations were conducted with $Re_{m}\approx 112$ and $Re^{i}\approx 426$. }
\label{fig:reconnection4}
\end{figure} 

\begin{figure}[h!t]
\includegraphics[trim=115 505 132 65, clip, width=5.5in]{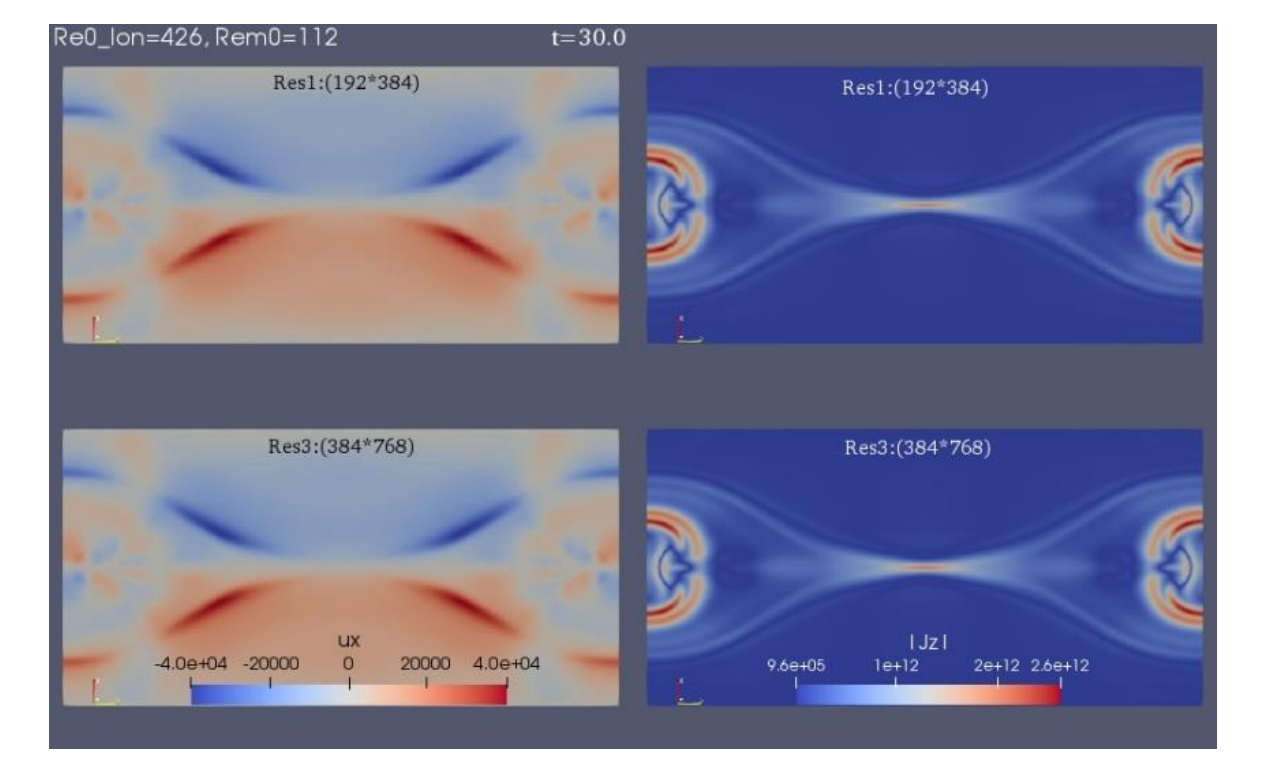}
\caption{The contours of vertical velocity, $V_{x}$, and spanwise current density, $\vert J_{z} \vert$, with different resolutions for $Re_{m}\approx 112$ and $Re^{i}\approx 426$.}
\label{fig:reconnection5}
\end{figure}

A grid convergence test has been conducted for the magnetic reconnection problem and the reconnection fluxes calculated using the two-fluid plasma solver are converged at a moderate grid resolution (e.g. $256\times512$) for $Re^{i}=426$ and $Re_{m}=112$, as shown in Fig.~\ref{fig:reconnection4}(a). By using the finest grid (e.g. $768\times1536$) results as the ‘exact’ solutions, one can calculate the numerical error on coarser grids. The results are shown in Fig.~\ref{fig:reconnection4}(b). The grid convergence rate estimated from the last two point is $\hat{n}=6.58$, which is fairly close to the theoretical limit of sixth-order compact scheme. 
Fully converged DNS-like results can also be observed at cases with higher viscous Reynolds provided the grid resolution is sufficiently large. For example, the CFDNS results for the case with initial ion Reynolds number $Re^{i}=1029$ (not shown) are fully converged at grid resolution $384\times768$. In this test case, the threshold value for fully converged DNS-like results is around $Re_\Delta\approx 2.6$. In addition to the reconnection flux profile, the 2D contours of vertical velocity and spanwise current density shown in Fig.~\ref{fig:reconnection5} further confirm that the CFDNS results presented here are indeed fully converged DNS-like solutions.

\begin{figure}[h!t]
\includegraphics[trim=75 522 80 70, clip, width=6.4in]{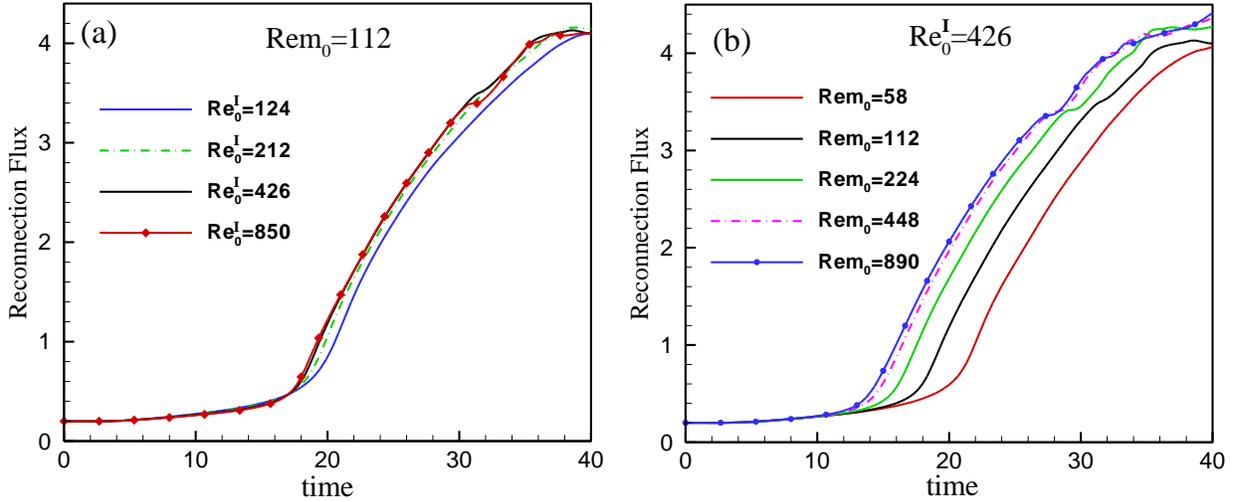}
\caption{The temporal variations of reconnection flux for different viscosities and magnetic resistivities.}
\label{fig:reconnection6}
\end{figure} 

Finally, to examine the effects of the plasma transport terms and convergence of the results with the Reynolds number, we have conducted a series of simulations for a range of $Re^{i}$ and $Re_{m}$ values. First, as observed in Fig.~\ref{fig:reconnection6}(a), viscosity has a slight delay effect on the reconnection time, as  $Re^{i}$ is increased from $124$ to $850$. However, for all viscous Reynolds numbers, $Re^{i}$, the magnetic flux saturates to the same non-dimensional value of around $4.0$ at a non-dimensional time close to $40.0$. These values are consistent with those reported in Refs.~\cite{Birn2001,Ma2001}. In addition, the time variations of the reconnection flux quickly converge at $Re^{i}$ values above $\sim400$. For $Re^{i}=426$, Fig.~\ref{fig:reconnection6}(b) shows that the magnetic flux also converges as the magnetic Reynolds number, is increased to $Re_m=890$. 

Based on this convergence, we assess that the results obtained with $Re^{i}=426$ and $Re_m=890$ fully represent the collisionless reconnection process for the number density and length scale shown above, representative of solar flare reconnection. In addition, due to the robustness of the saturation flux value and time to the viscosity value, the results are also very similar to numerical results relying on numerical  dissipation for regularization. On the other hand, resistivity has a larger effect than viscosity at moderate  $Re_m$ values (Fig.~\ref{fig:reconnection6}b), in particular on the reconnection time. Lower resistivity or large $Re_{m}$ values lead to an earlier reconnection and slightly larger saturation flux values. The reconnection time and saturation flux value for the case $Re_{m}=224$ shown in Fig.~\ref{fig:reconnection6}(b) are very close to those predicted by previous Hall-MHD simulations~\cite{Toth2008} with $Re_{m}=200$. However, due to the larger potential effect of the numerical regularization scheme regarding resistivity, predictions using the ideal (inviscid and perfect conductivity) two-fluid plasma model (e.g. Ref.~\cite{Hakim2006}), need to be evaluated with more caution. 
However, again, the convergence of the results as $Re^i$ and $Re_m$ are increased, explains why previous studies using relatively low $Re_{m}$ values, in comparison to the extremely high $Re_{m}$ values found in practice, are still useful in predicting the magnetic reconnection phenomena occurring in space.      

\section{CONCLUSION}
\label{sec:5}

In this study, to be able to generate high-order fully converged DNS-like solutions for plasma flow problems, we have implemented the Braginskii two-fluid plasma model with full plasma transport terms, including temperature and magnetic field dependent ion and electron viscous stresses and heat fluxes, frictional drag force, and ohmic heating term, in the CFDNS code, using sixth-order non-dissipative compact finite differences with negligible numerical dissipation/diffusion. To maintain computational feasibility, while also solving all the dynamically relevant time and length scales, the infinite speed of light and negligible electron inertia assumptions have been used. 

The range of applicability of the resulting two-fluid plasma equations is discussed in detail.  
This was achieved by using a non-dimensional analysis of the equations, which highlights the relevant non-dimensional parameters. These parameters are cast in terms of characteristic scales found in practical problems of interests, including the characteristic number density, $n_0$, length cle $L_0$, temperature, $T_0$, and magnetic strength, $B_0$. The non-dimensional parameters can be used to estimate the relative contributions of Hall and Biermann battery effects, resistivity, viscous stress, and heat flux in different regimes. In the appropriate limits, the two-fluid plasma equations recover the conventional MHD (i.e. ideal, resistive, and Hall) equations. First, the corresponding non-dimensional single fluid equations for the mixture velocity, density, pressure and temperature are derived. Then, (i) conventional Hall-MHD equations can be recovered in regimes where $0\ll\hat{\lambda}_{i}\ll\sqrt{m_{i}/m_{e}}$ (and/or $0\ll\hat{r}_{Li}\ll\sqrt{m_{i}/m_{e}}$), $Re^{i},\ Re^{e}\rightarrow\infty$, and $Fr\rightarrow\infty$; (ii) Resistive MHD equations can be recovered in regimes where $\hat{\lambda}_{i}\rightarrow0$ (and/or $\hat{r}_{Li}\rightarrow0$), $Fr\rightarrow\infty$, and $Re^{i},\  Re^{e}\rightarrow\infty$, but $Re_{m}$ has a finite value; and (iii) Ideal MHD equations are recovered in regimes when in addition $Re_{m}\rightarrow\infty$. The single fluid, as well as the conventional Hall-MHD equations, are unclosed due to the explicit presence of the electron pressure. Several choices for closing these equations in previous studies are discussed. In contrast, the two-fluid plasma equations are more general and do not need additional assumptions.

The two-fluid solver was demonstrated against four canonical problems, to confirm its accuracy and robustness in handling plasma flows in different regimes. These test cases include the Alfven and whistler waves for parameter values relevant to solar corona, the electromagnetic Brio-Wu plasma shock~\cite{Brio1988} with parameter values relevant to the bow shock caused by the interaction between solar wind and earth’s magnetosphere, and the fast magnetic reconnection in solar flares. All physical transport terms are retained for the four test cases, and the convergence with respect to the viscous and magnetic Reynolds numbers was discussed, in addition to proving grid convergence of the results. For both Alfven and whistler dispersion relations the numerical results are in excellent agreement with the analytical or linear stability theory (LST) predictions for corresponding ideal MHD and Hall-MHD equations over a wide range of wavenumbers. Because of the inclusion of physical viscosity in the two-fluid plasma solver, all plasma shock characteristics can be fully resolved at all Reynolds numbers, provided the grid resolution is sufficiently high. This means that the the ion grid Reynolds number, $Re_{\Delta}=Re^{i}/NX$ needs to be smaller than a threshold value, which for the plasma shock test case is around $2.3$. Near the sharp gradients in the plasma shock problem, in contrast to the first-order convergence rate commonly found in studies using shock-capturing schemes, the grid convergence rate calculated here is in the range of $\hat{n}\sim6.08-6.14$ which is very close to the theoretical value of the sixth-order compact scheme. 

For the last test case, the CFDNS results successfully demonstrate, using the two-fluid plasma model, the fast magnetic reconnection process occurring under solar flare conditions. The magnetic flux saturation time and value predicted here are in good agreement with those reported in previous studies under similar conditions. The systematic examination of $Re^{i}$ and $Re_{m}$ effects on the magnetic reconnection reveals that the results are converged for the largest values used in this study, $Re^{i}=426$ and $Re_{m}=890$. This implies that the results are relevant to practical problems with much larger Reynolds numbers. The viscous effects are relatively small for $Re^i\sim100-400$, so that 
coarse resolution simulation results using numerical dissipation to regularize the equations are likely close to the high Reynolds number results. On the other hand, the reconnection flux saturation value and time are more sensitive to changes in $Re_{m}$. The CFDNS results with $Re_m\sim200$ are close to those reported in previous studies, but the results become converged at much higher magnetic Reynolds number values ($Re_m>800$).  These results are particularly useful in evaluating the different approximations used in plasma solvers (e.g. with/without viscosity, heat flux, resistivity, etc.). 

In general, the Braginskii transport coefficients become inaccurate for degenerate and/or partially ionized plasmas. However, more general formulations do not include full directional dependence of the physical transport with respect to the magnetic field or are less accurate for low-Z materials. Future simulations will address the importance of anisotropic transport, differences with more accurate models where available (e.g. for higher-Z materials), and further explore the existence of a mixing transition in various applications.

\begin{acknowledgments}
This work was made possible in part by funding from the LDRD program at Los Alamos National Laboratory through project number 20150568ER. Z. Li was supported by Los Alamos National Laboratory, under Grant No. 430461. Los Alamos National Laboratory is operated by Triad National Security, LLC, for the National Nuclear Security Administration of U.S. Department of Energy (Contract No. 89233218CNA000001). Computational resources were provided by the Los Alamos National Laboratory Institutional Computing Program and the High Performance Computing Center at Texas A$\&$M University-Corpus Christi.
\end{acknowledgments}

\appendix
\section{Plasma Transport Term Formulations}
\label{sec:A}
For completeness, the formulations for all transport terms, mostly following Ref.~\cite{braginskii1965}, as well as details of their implementation are given below. 

\subsection{Viscous stress tensors, $\mathbf{\pi}_{s}$}
\label{sec:viscous}
In general, three major steps are needed for calculating the viscous stress. First, the strain rate tensor, $\textbf{W}_{s}$, is calculated in the fixed Cartesian coordinate system,$\lbrace \mathbf{e}_1, \mathbf{e}_2, \mathbf{e}_3 \rbrace$, as: 
\begin{equation}\label{eq:A1}
\textbf{W}_{s}=-\left[\nabla\textbf{u}_{s}+\left(\nabla\textbf{u}_{s}\right)^{T}-\frac{2}{3}\left(\nabla\cdot\textbf{u}_{s}\right) \mathbf{I}\right] 
\end{equation}

\noindent
where $\mathbf{I}$ is the second-order identity tensor.

The next step is to restate the strain rate tensor, $\textbf{W}_{s}$, into a moving coordinate system aligned with the magnetic field, $\lbrace \mathbf{e}_1^{'}, \mathbf{e}_2^{'}, \mathbf{e}_3^{'} \rbrace$, in which $\mathbf{e}_3^{'}=\textbf{B}/\vert\textbf{B}\vert$ denotes the unity vector in the direction of magnetic field, as given below: 
\begin{equation}\label{eq:A2}
\textbf{W}_{s}^{'}=\mathbf{Q}^{T}\textbf{W}_{s}\mathbf{Q}.
\end{equation}

\noindent
The transformation matrix, $\mathbf{Q}$, is defined by: 

\begin{equation}\label{eq:A3}
\mathbf{Q}
=
\begin{bmatrix}
-B_{2}^{'} & -B_{1}^{'}B_{3}^{"} & B_{1}^{"} \\
 B_{1}^{'} & -B_{2}^{'}B_{3}^{"} & B_{2}^{"} \\
 0 & B_{1}^{'}B_{1}^{"}+B_{2}^{'}B_{2}^{"} & B_{3}^{"}
\end{bmatrix}
\end{equation}

\noindent where $B_{i}^{'}=B_{i}/\sqrt{B_1^2+B_2^2}$ and $B_{i}^{"}=B_{i}/\sqrt{B_1^2+B_2^2+B_3^2}$.
The viscous stress in the new coordinate system can then be calculated as:

\begin{eqnarray} 
\label{eq:A4b}
\pi_{11,s}^{'}&=&-\frac{1}{2}\mu_{0}^{s}\left(W_{11,s}^{'}+W_{22,s}^{'}\right)-\frac{1}{2}\mu_{1}^{s}\left(W_{11,s}^{'}-W_{22,s}^{'}\right)-\mu_{3}^{s}W_{12,s}^{'}\ ,\\
\label{eq:A4d}
\pi_{12,s}^{'}&=&\pi_{21,s}^{'}=-\mu_{1}^{s}W_{12,s}^{'}+\frac{1}{2}\mu_{3}^{s}\left(W_{11,s}^{'}-W_{22,s}^{'}\right), \\ 
\label{eq:A4e}
\pi_{13,s}^{'}&=&\pi_{31,s}^{'}=-\mu_{2}^{s}W_{13,s}^{'}-\mu_{4}^{s}W_{23,s}^{'}\ ,\\
\label{eq:A4c}
\pi_{22,s}^{'}&=&-\frac{1}{2}\mu_{0}^{s}\left(W_{11,s}^{'}+W_{22,s}^{'}\right)-\frac{1}{2}\mu_{1}^{s}\left(W_{22,s}^{'}-W_{11,s}^{'}\right)+\mu_{3}^{s}W_{12,s}^{'}\ ,\\
\label{eq:A4f}
\pi_{23,s}^{'}&=&\pi_{32,s}^{'}=-\mu_{2}^{s}W_{23,s}^{'}+\mu_{4}^{s}W_{13,s}^{'}\ ,\\
\label{eq:A4a}
\pi_{33,s}^{'}&=&-\mu_{0}^{s}W_{33,s}^{'}\ ,
\end{eqnarray}

\noindent
where $\mu_{j}^{s}$, $j=1,..,4$, are the ion and electron viscosity coefficients which are mainly functions of temperature, $T_s$, and number density, $n_s$. For ions, one has $\mu_{0}^{i}=\left(2.23/2.33\right)n_{i}k_{B}T_{i}\tau_{i}$, $\mu_{2}^{i}=n_{i}k_{B}T_{i}\tau_{i}\left(1.2x^2+2.23\right)/\Delta$, $\mu_{4}^{i}=n_{i}k_{B}T_{i}\tau_{i}x\left(x^2+2.38\right)/\Delta$, where $x=\omega_{ci}\tau_{i}$ and $\Delta=x^{4}+4.03x^{2}+2.33$. The coefficients $\mu_{1}^{i}$ and $\mu_{3}^{i}$ can be obtained by replacing $x$ by $2x$ in the formulations for $\mu_{2}^{i}$ and $\mu_{4}^{i}$, respectively. Here, $\tau_{i}=12\pi^{3/2}\varepsilon_{0}^{2}\sqrt{m_{i}}\left(k_{B}T_{i}\right)^{3/2}/\left(ln\Lambda e^{4}Z^{4}n_{i}\right)$ is the ion collision time and $\omega_{ci}=Ze\vert\textbf{B}\vert/m_{i}$ is the ion cyclotron frequency.  For electrons, one has $\mu_{0}^{e}=\left(8.50/11.6\right)n_{e}k_{B}T_{e}\tau_{e}$, $\mu_{2}^{e}=n_{e}k_{B}T_{e}\tau_{e}\left(2.05x^2+8.50\right)/\Delta$, $\mu_{4}^{e}=-n_{e}k_{B}T_{e}\tau_{e}x\left(x^2+7.91\right)/\Delta$, where $x=\omega_{ce}\tau_{e}$ and $\Delta=x^{4}+13.8x^{2}+11.6$. Similarly, the coefficients $\mu_{1}^{e}$ and $\mu_{3}^{e}$ can be obtained by replacing $x$ by $2x$ in the formulations for $\mu_{2}^{e}$ and $\mu_{4}^{e}$, respectively. Here, $\tau_{e}=6\sqrt{2}\pi^{3/2}\varepsilon_{0}^{2}\sqrt{m_{e}}\left(k_{B}T_{e}\right)^{3/2}/\left(ln\Lambda e^{4}Zn_{e}\right)$ is the electron collision time and $\omega_{ce}=e\vert\textbf{B}\vert/m_{e}$ is the electron cyclotron frequency.

In this study, the Coulomb logarithm formula, $ln\Lambda$, is adopted from Ref.~\cite{Murillo2016} and its expression in Gaussian units is given below:  
\begin{equation}\label{eq:A5}
ln\Lambda=
\begin{cases}
-ln\left(\sum_{k=1}^{5} a_{k}g^{k} \right) 
  , & \text{if} \quad g=(Ze)^2/k_B\lambda_{eff}T_{e}\leqslant 1  \\
2\times \dfrac{b_0+b_1ln(g)+b_2ln^2(g)}{1+b_3g+b_{4}g^2}, & \text{if} \quad g=(Ze)^2/k_B\lambda_{eff}T_{e}>1
\end{cases}
\end{equation}
\noindent
The numerical values of the constant coefficients $a_{1},a_{2},...,a_{5}, b_{0},...,b_{4}$ can be found in Ref.~\cite{Murillo2016}. The effective screening length $\lambda_{eff}$ can be estimated as:
\begin{equation}\label{eq:A51}
\lambda_{eff}=\lambda_{e}\left(1+\dfrac{1}{1+3\Gamma}\right)^{-1/2} 
\end{equation}

\noindent
where $\lambda_{e}=\left[k_BT_{e}/(4\pi Z^2e^2n_e)\right]^{1/2} $, $\Gamma=(Ze)^2/\hat{a}_{i}k_BT_{e}$, and $\hat{a}_{i}=\left(3/4\pi n_e \right)^{1/3}$.

Finally, the viscous stress tensor, $\mathbf{\pi}_{s}$, can be obtained by restating $\mathbf{\pi}_{s}^{'}$ back into the fixed coordinate system, $\lbrace \mathbf{e}_1, \mathbf{e}_2, \mathbf{e}_3 \rbrace$, as shown below:
\begin{equation}\label{eq:A6}
\mathbf{\pi}_{s}=\mathbf{Q}\mathbf{\pi}_{s}^{'}\mathbf{Q}^{T} 
\end{equation}

For the special case without magnetic field, i.e. $\textbf{B}=0$, the viscous stress tensor can be calculated directly by using the following formulation, 
\begin{equation}\label{eq:A7}
\mathbf{\pi}_{s}=\mu_{0}^{s}\textbf{W}_{s} 
\end{equation}

Another special situation is when the magnetic field is aligned with the fixed coordinate system,  i.e. $B_{1}=B_{2}=0$ and $B_{3}\neq 0$.  In this case, the transformation matrix, $\mathbf{Q}$, is reduced to the second-order identity tensor $\mathbf{I}$. Therefore, no coordinate transformation is needed and the viscous stress tensor can be calculated by using equations (\ref{eq:A4b})-(\ref{eq:A4a}) directly.  

\subsection{Heat Flux, $\mathbf{q}_{s}$}
\label{sec:heatflux}

The ion heat flux, $\mathbf{q}_{i}$, is caused by temperature gradient only and can be expressed as:
\begin{equation}\label{eq:A8}
\mathbf{q}_{i}=-\left(\frac{n_{i}k_{B}^{2}T_{i}\tau_{i}}{m_{i}}\right)\left[\eta_{0}\nabla_{\parallel}T_{i} + \left(\frac{\eta_{1}^{'}x^{2}+\eta_{0}^{'}}{\Delta}\right) \nabla_\perp T_{i} - \frac{x\left(\eta_{1}^{"}x^{2}+\eta_{0}^{"}\right)}{\Delta}\left(\textbf{h}\times\nabla T_i \right)\right],
\end{equation} 

\noindent
where $\textbf{h}=\textbf{B}/\vert\textbf{B}\vert$ represents a unity vector in the direction of local magnetic field and the symbols $\parallel$ and $\perp$ on any vector denote its component in the parallel or perpendicular direction to the magnetic field, $\textbf{B}$, respectively. For example, $\nabla_{\parallel}T_{i}=\textbf{h}\left(\textbf{h}\cdot\nabla T_i\right)$ and $\nabla_\perp T_{i}=\textbf{h}\times\left(\nabla T_i\times\textbf{h}\right)=\nabla T_i-\nabla_{\parallel}T_{i}$. The non-dimensional variables $x$ and $\Delta$ follow the definitions above.

On the contrary, the electron heat flux, $\mathbf{q}_{e}$, is caused by both temperature gradient and the relative velocity between ion and electron, $\left(\textbf{u}_{i}-\textbf{u}_{e}\right)$ or current density, $\textbf{J}$, and can be written as $\mathbf{q}_{e}=\mathbf{q}_{Te}+\mathbf{q}_{ue}$. The two parts are formulated as:
\begin{eqnarray}
\label{eq:A9}
\mathbf{q}_{Te}&=&-\left(\frac{n_{e}k_{B}^{2}T_{e}\tau_{e}}{m_{e}}\right)\left[\gamma_{0}\nabla_{\parallel}T_{e} + \left(\frac{\gamma_{1}^{'}x^{2}+\gamma_{0}^{'}}{\Delta}\right) \nabla_\perp T_{e} - \frac{x\left(\gamma_{1}^{"}x^{2}+\gamma_{0}^{"}\right)}{\Delta}\left(\textbf{h}\times\nabla T_e \right)\right],\\
\label{eq:A10}
\mathbf{q}_{ue}&=&-\frac{k_{B}T_{e}}{e} \left[\beta_{0}\textbf{J}_{\parallel} + \left(\frac{\beta_{1}^{'}x^{2}+\beta_{0}^{'}}{\Delta}\right) \textbf{J}_\perp  + \frac{x\left(\beta_{1}^{"}x^{2}+\beta_{0}^{"}\right)}{\Delta}\left(\textbf{h}\times\textbf{J} \right)\right].
\end{eqnarray} 

\noindent
The numerical values of the constant coefficients, $\eta_{0}$, $\gamma_{0}$, $\beta_{0}$, $\gamma_{0}^{'}$, etc., can be found in Ref.~\cite{braginskii1965}.

\subsection{Frictional drag force, $\textbf{R}_{s}$}
\label{sec:dragforce}

Similar to the electron heat flux, $\mathbf{q}_{e}$, the frictional drag force between ions and electrons, $\textbf{R}_{ei}$ (or $\textbf{R}_{e}$), also has two different contributions:

\begin{eqnarray}
\label{eq:A11}
\textbf{R}_{u}&=&\left(\frac{m_{e}}{e\tau_{e}}\right) \left[\alpha_{0}\textbf{J}_{\parallel} + \left(1-\frac{\alpha_{1}^{'}x^{2}+\alpha_{0}^{'}}{\Delta}\right) \textbf{J}_\perp  - \frac{x\left(\alpha_{1}^{"}x^{2}+\alpha_{0}^{"}\right)}{\Delta}\left(\textbf{h}\times\textbf{J} \right)\right],\\ 
\label{eq:A12}
\textbf{R}_{T}&=&-\left(n_{e}k_{B}\right)\left[\beta_{0}\nabla_{\parallel}T_{e} + \left(\frac{\beta_{1}^{'}x^{2}+\beta_{0}^{'}}{\Delta}\right) \nabla_\perp T_{e} + \frac{x\left(\beta_{1}^{"}x^{2}+\beta_{0}^{"}\right)}{\Delta}\left(\textbf{h}\times\nabla T_e \right)\right].
\end{eqnarray} 

\noindent
$\textbf{R}_{u}$ is the classical momentum frictional force caused by the velocity difference between ions and electrons, while the thermal frictional force, $\textbf{R}_{T}$, is produced by the electron temperature gradient. 

\subsection{Collision generated heat, $Q_{s}$}
\label{sec:collisionheat}

Following the approximations made in Refs.~\cite{braginskii1965, Callen2006}, the ion and electron collision generated heat terms are written as:
\begin{equation}\label{eq:A13}
Q_i=Q_{\Delta}=3\left(\frac{m_e}{m_i}\right)\left(\frac{n_e}{\tau_{e}}\right)k_{B}\left(T_e -T_i\right)  
\end{equation} 
\begin{equation}\label{eq:A14}
Q_e=\textbf{R}_{e}\cdot\left(\textbf{u}_{i}-\textbf{u}_{e}\right)-Q_{\Delta}   
\end{equation} 

\noindent
The expression $\textbf{R}_{u}\cdot\left(\textbf{u}_{i}-\textbf{u}_{e}\right)$ is the general ohmic heating term.

We note here that the Braginskii coefficients are consistently derived using two-term Sonine polynomial solutions of the Boltzmann equation. The electron conductivity model presented in Ref.~\cite{Lee-More1984} reduces to the Braginskii model for fully ionized nondegenerate plasmas, but retains a higher precision for the numerical coefficients due a different treatment of the Sonine polynomial solution. Thus, most of the coefficients appearing in the heat flux and frictional drag force Braginskii formulas are about $8\%$ different than the exact values. On the other hand, Ref.~\cite{Lee-More1984} does not include electron-electron scattering, which ads a non-negligible contribution for low-Z materials, where the model overestimates the conductivity. For consistency with the other transport formulas and to consider the full directional dependence of the transport, here, we use the Braginskii formulation for the heat flux and frictional drag force and will address the differences compared to Ref.~\cite{Lee-More1984} formulation elsewhere.

\section{Quasi-neutrality condition}
\label{sec:neutrality}

As discussed in the Section \ref{sec:finaleq}, an indication of the accuracy of the numerical integration is that the charge density, $\rho_{c}$, evaluated from the divergence of electric field, \textbf{E}, remains sufficiently small at all the times. For all test cases discussed in this paper, the maximum normalized charge density in the computational domain, $\vert \overline{\rho_{c}}\vert_{max}=\vert \varepsilon_{0}\nabla\cdot\textbf{E}/\left(eZn_i\right)\vert_{max}$, was monitored throughout the  simulation times. 

\begin{figure}[h!b]
\includegraphics[trim=70 522 75 75, clip, width=6.4in]{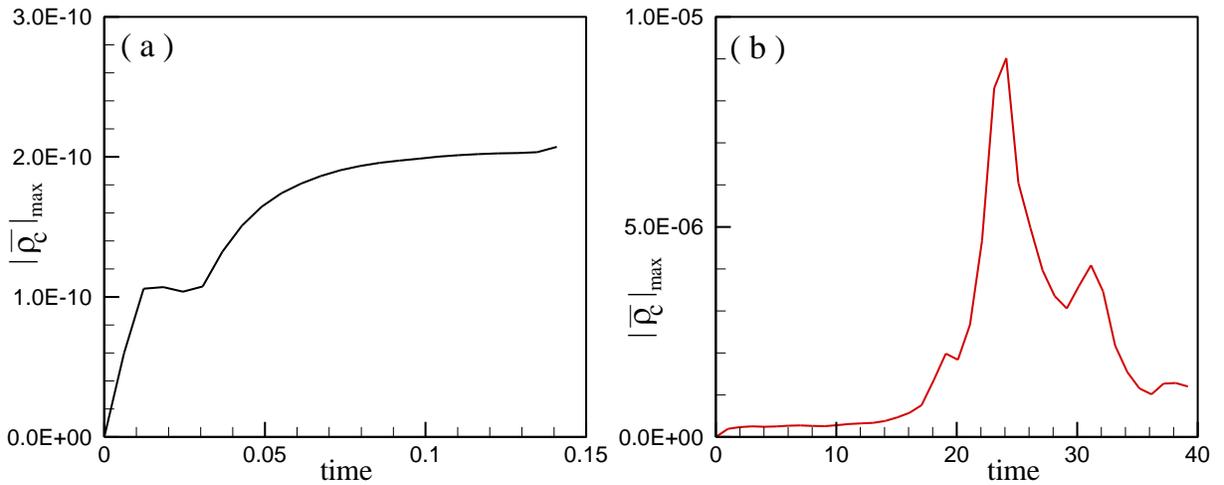}
\caption{The temporal variations of maximum normalized charge density for (a) plasma shock and (b) magnetic reconnection cases.}
\label{fig:appendix1}
\end{figure} 

Fig.~\ref{fig:appendix1} shows $\vert \overline{\rho_{c}}\vert_{max}$ variation for the 1D plasma shock and 2D magnetic reconnection problems. Both cases exhibit sufficiently small values to conclude that the simulations conducted in this study satisfy the quasi-neutrality condition.  

\section{Single-fluid limiting equations}
\label{sec:limiting}

The single-fluid plasma equations (\ref{eq:28})-(\ref{eq:31}) described in Section \ref{sec:OneND} are derived from the non-dimensional two-fluid plasma equations (\ref{eq:20})-(\ref{eq:27}) by using the ion-electron mixture definitions, infinite speed of light, and negligible electron inertia assumptions. The Hall term, electron pressure term, and all plasma transport terms are retained in the single-fluid plasma equations, which can be viewed as full Hall-MHD equations, in contrast to the conventional Hall-MHD equations where the viscous, heat flux and acceleration terms are neglected. As shown in this Appendix, the conventional Hall, resistive, and ideal MHD equations can be recovered from the general single-fluid equations (\ref{eq:28})-(\ref{eq:31}) as limiting cases.   

\subsection{The conventional Hall-MHD equations}
\label{sec:Hall-MHD}
In regimes where $Re^{i}, Re^{e}\rightarrow\infty$, $\textbf{q}_{ue}^{*} \rightarrow\ 0$, and $\textbf{R}_{T}^{*} \rightarrow\ 0$, and assuming that the gradients stay finite, the single-fluid equations (\ref{eq:28})-(\ref{eq:31}) given in Section \ref{sec:OneND} reduce to:  
\begin{eqnarray}
\label{eq:C1}
\pD{\rho^{*}}{t^{*}}+\nabla^{*}\cdot\left(\rho^{*}\textbf{u}^{*}\right)&=&0,\\
\label{eq:C2}
\pD{\left(\rho^{*}\textbf{u}^{*}\right)}{t^{*}}+\nabla^{*}\cdot\left(\rho^{*}\textbf{u}^{*}\textbf{u}^{*}\right)&=&-\beta\nabla^{*}p^{*}+\textbf{J}^{*}\times\textbf{B}^{*}, \\                               
\label{eq:C3}
\frac{1}{\gamma-1}\left[\pD{p^{*}}{t^{*}}+\nabla^{*}\cdot\left(p^{*}\textbf{u}^{*}\right)   \right]&=&-p^{*}\nabla^{*}\cdot\textbf{u}^{*}+\frac{\hat{\lambda}_{i}}{\gamma-1}\nabla^{*}\cdot\left(\frac{p_{e}^{*}}{\rho^{*}}\textbf{J}^{*}\right)+\hat{\lambda}_{i}p_{e}^{*}\nabla^{*}\cdot\left(\frac{\textbf{J}^{*}}{\rho^{*}}\right)+ 
\protect \nonumber \\ &&\frac{1}{\beta Re_{m}}\textbf{R}_{\text{u}}^{*}\cdot\textbf{J}^{*} \\
\label{eq:C4}
\pD{\textbf{B}^{*}}{t^{*}}&=&-\nabla^{*}\times\textbf{E}^{*},\\
\label{eq:C5}
\textbf{E}^{*}+\textbf{u}^{*}\times\textbf{B}^{*}&=&\hat{\lambda}_{i}\frac{1}{\rho^{*}}\textbf{J}^{*}\times\textbf{B}^{*}-\hat{\lambda}_{i}\frac{\beta}{\rho^{*}}\nabla^{*}p_{e}^{*}+\frac{1}{Re_{m}}\textbf{R}_{\text{u}}^{*}\\
\label{eq:C6}
\textbf{J}^{*}&=&\nabla^{*}\times\textbf{B}^{*}.
\end{eqnarray}
\noindent
Equations (\ref{eq:C1})-(\ref{eq:C6}) are the conventional Hall-MHD equations \cite{Huba2003,Callen2006} in which $\textbf{R}_{\text{u}}$ is function of current density, $\textbf{J}^{*}$, as shown in Appendix \ref{sec:dragforce} and $\frac{1}{\beta Re_{m}}\textbf{R}_{\text{u}}^{*}\cdot\textbf{J}^{*}\propto \frac{1}{\beta Re_{m}}\textbf{J}^{*}\cdot\textbf{J}^{*}$ represents the resistive effects. The $\textbf{q}_{ue}^{*}$ and $\textbf{R}_{T}^{*}$ terms have been never been considered in previous derivations of Hall-MHD equations. For the tests considered in this study, these two terms are negligible. In the presence of turbulence, it is assumed that viscous dissipation does not vanish in the infinite Reynolds number limit, so that the domain of applicability of equations (\ref{eq:C1})-(\ref{eq:C6}) relates to non-turbulent flows, unless they are used in the context of turbulence modeling with added subgrid or turbulent transport models.

Written as above, the conventional Hall-MHD equations are not closed, due to the presence of the electron pressure, $p_{e}^{*}$, which cannot be estimated from the rest of the variables. In practice, to close the equations, some studies~\cite{Callen2006,Dmitruk2006,Toth2008} simply neglect the electron pressure, while others~\cite{Browning2014,Srinivasan2011} assume identical ion and electron temperatures, $T_{i}^{*}= T_{e}^{*}$. In the latter case, the electron pressure becomes $p_{e}^{*}=Z p_{i}^{*}=p^*/\left(1+1/Z\right)$. 
   
\subsection{The resistive MHD equations}
\label{sec:resistive-MHD}
In regimes where  $\hat{\lambda}_{i}\rightarrow0$ (and/or $\hat{r}_{Li}\rightarrow0$), $Re^{i}, Re^{e}\rightarrow\infty$, and $Fr\rightarrow\infty$, the single-fluid equations (\ref{eq:28})-(\ref{eq:31}) reduce to:  
\begin{eqnarray}
\label{eq:C7}
\pD{\rho^{*}}{t^{*}}+\nabla^{*}\cdot\left(\rho^{*}\textbf{u}^{*}\right)&=&0, \\
\label{eq:C8}
\pD{\left(\rho^{*}\textbf{u}^{*}\right)}{t^{*}}+\nabla^{*}\cdot\left(\rho^{*}\textbf{u}^{*}\textbf{u}^{*}\right)&=&-\beta\nabla^{*}p^{*}+\textbf{J}^{*}\times\textbf{B}^{*}, \\                             
\label{eq:C9}
\frac{1}{\gamma-1}\left[\pD{p^{*}}{t^{*}}+\nabla^{*}\cdot\left(p^{*}\textbf{u}^{*}\right)   \right]&=&-p^{*}\nabla^{*}\cdot\textbf{u}^{*}+ \frac{1}{\beta Re_{m}}\textbf{R}_{\text{u}}^{*}\cdot\textbf{J}^{*} \\
\label{eq:C10}
\pD{\textbf{B}^{*}}{t^{*}}&=&-\nabla^{*}\times\textbf{E}^{*},\\
\label{eq:C11}
\textbf{E}^{*}+\textbf{u}^{*}\times\textbf{B}^{*}&=&\frac{1}{Re_{m}}\textbf{R}_{\text{u}}^{*},\\
\label{eq:C12}
\textbf{J}^{*}&=&\nabla^{*}\times\textbf{B}^{*}
\end{eqnarray}
\noindent
Equations (\ref{eq:C7})-(\ref{eq:C12}) are the non-dimensional resistive MHD equations \cite{Freidberg1982,Callen2006}. Obviously, the resistive HMD equations are closed without the need of explicitly assuming identical ion and electron temperatures ($T_{i}^{*}= T_{e}^{*}$). As before, $\textbf{R}_{\text{u}}$ is function of current density, $\textbf{J}^{*}$. Again, neglecting the viscous contributions in the infinite Reynolds number limit, generally precludes the use of equations (\ref{eq:C7})-(\ref{eq:C12}) for turbulent flow calculations.

\subsection{The ideal MHD equations}
\label{sec:ideal-MHD}
If in addition, $Re_{m}\rightarrow\infty$, the resistive MHD equations (\ref{eq:C7})-(\ref{eq:C12}) can be further reduced to the ideal-MHD equations \cite{Freidberg1982,Callen2006} given below:
\begin{eqnarray}
\label{eq:C13}
\pD{\rho^{*}}{t^{*}}+\nabla^{*}\cdot\left(\rho^{*}\textbf{u}^{*}\right)&=&0,\\
\label{eq:C14}
\pD{\left(\rho^{*}\textbf{u}^{*}\right)}{t^{*}}+\nabla^{*}\cdot\left(\rho^{*}\textbf{u}^{*}\textbf{u}^{*}\right)&=&-\beta\nabla^{*}p^{*}+\textbf{J}^{*}\times\textbf{B}^{*}, \\                          
\label{eq:C15}
\frac{1}{\gamma-1}\left[\pD{p^{*}}{t^{*}}+\nabla^{*}\cdot\left(p^{*}\textbf{u}^{*}\right)   \right]&=&-p^{*}\nabla^{*}\cdot\textbf{u}^{*},\\
\label{eq:C16}
\pD{\textbf{B}^{*}}{t^{*}}&=&-\nabla^{*}\times\textbf{E}^{*},\\
\label{eq:C17}
\textbf{E}^{*}+\textbf{u}^{*}\times\textbf{B}^{*}&=&0,\\
\label{eq:C18}
\textbf{J}^{*}&=&\nabla^{*}\times\textbf{B}^{*}.
\end{eqnarray}
\noindent

\bibliography{RT_Plasma_refs}   

\end{document}